\documentclass[a4paper,11pt]{article}
\usepackage{jheppub} 
\makeatletter
\gdef\@fpheader{\textcolor{white}{preprint}}
\makeatother
\usepackage{lineno}
\usepackage{comment}

\title{\boldmath Matrix Bootstrap Approximation without Positivity Constraint}

\author{Reishi Maeta}
\affiliation{
Graduate School of Advanced Science and Engineering,\\
Hiroshima University,\\
Higashi-Hiroshima, Hiroshima 739-8526, Japan
}
\affiliation{
Department of Physics,\\
McGill University,\\
Montreal, QC H3A 2T8, Canada
}

\emailAdd{maeta-reishi@hiroshima-u.ac.jp}

\abstract{
We propose a bootstrap approximation method for the Hermitian one-matrix model that does not rely on positivity constraints.
The theoretical foundation of this method is that the one-matrix model admits an eigenvalue distribution $\rho(\lambda)$, and that the moments $w_n$ generated from it satisfy the loop equations.
Our framework is designed to numerically determine a self-consistent pair of $\rho(\lambda)$ and $w_n$ that simultaneously satisfies these two requirements.
In the concrete implementation the least-squares method is employed, and since the sign problem is absent in this formulation, the method can be formally applied to the Minkowski one-matrix model as well, provided that the one-cut structure of the resolvent is assumed.
Actual numerical calculations show that this bootstrap approximation reproduces, with very high accuracy, the exact solutions for Euclidean-type models and the perturbative results for Minkowski-type models.}

\begin{document}
\maketitle
\flushbottom

\section{Introduction}
\label{sec:Introduction}

Matrix models continue to be an important subject of research today, attracting sustained interest from a wide range of perspectives, including string theory, low-dimensional quantum gravity, and (lattice) gauge theory. Among them, \textit{large-$N$ matrix models} occupy a particularly prominent position. Starting from large-$N$ reduced lattice gauge theories, famously known as the Eguchi--Kawai model~\cite{Eguchi:1982nm}, through Hermitian matrix models as nonperturbative formulations of Liouville theory, and further to the IKKT and BFSS matrix models~\cite{Ishibashi:1996xs, deWit:1988wri, Banks:1996vh} that are widely regarded as nonperturbative definitions of superstring theory, all of these theories remain active areas of intensive research. In addition, JT gravity~\cite{Jackiw:1984je, Teitelboim:1983ux, Almheiri:2014cka}, which has attracted considerable attention in recent years in connection with the holographic principle, can also be viewed as a matrix model~\cite{Saad:2019lba}, and it deserves special mention as a theory that symbolizes recent progress in our understanding of quantum gravity.

In parallel with analytical studies, matrix models have also been extensively investigated over many years by numerical simulations, mainly based on Monte Carlo methods~\cite{Kabat:2000zv, Ambjorn:2000dx, Martin:2004un, Azuma:2004zq, Panero:2006bx, Hanada:2008gy, Gonzalez-Arroyo:2010omx, Hanada:2011fq, Kim:2011cr, Gonzalez-Arroyo:2014dua, Filev:2015hia}. In matrix models, expectation values of physical observables are defined as statistical averages over matrix integrals, and such averages are highly suitable for direct numerical computation on a computer if $N$ is finite. Since analytical calculations for theories with finite \(N\) generally involve severe difficulties, the ability of Monte Carlo methods to analyze such theories non-perturbatively constitutes an enormous advantage.

On the other hand, as the matrix size \(N\) increases, direct evaluation of matrix integrals becomes progressively more difficult, and in the limit \(N \to \infty\) it becomes completely intractable.
\footnote{
It should be noted, however, that in order to probe the behavior of a theory in the limit $N \to \infty$, it is not always necessary to literally take $N$ to infinity.
For example, in Monte Carlo simulations of the TEK model, it has been repeatedly confirmed that large-$N$ reduction is reproduced even at a finite but sufficiently large value of $N$~\cite{Gonzalez-Arroyo:2010omx, Gonzalez-Arroyo:2014dua}.
A genuine necessity to take the limit $N \to \infty$ arises only in theories for which the qualitative behavior differs between finite $N$ and infinite $N$, namely when the $N \to \infty$ limit is (potentially) non-smooth.
}
A method well suited to such large-\(N\) matrix models is the \textit{matrix bootstrap} approach.
The earliest proposal which combined loop equations with positivity constraints was about lattice gauge theories~\cite{Anderson:2016rcw}, and this framework was subsequently applied to large-\(N\) matrix models~\cite{Lin:2020mme, Kazakov:2021lel}. Furthermore, methods for applying this framework to the BFSS matrix model are currently being developed~\cite{Han:2020bkb, Lin:2023owt, Lin:2024vvg}. On the other hand, the application of bootstrap methods to the IKKT matrix model has so far remained relatively unexplored. 
Nevertheless, it is known that this model admits several saddle-point solutions that are meaningful only in the large-$N$ limit, such as those characterized by the Moyal--Weyl--type commutation relations $[A^{\mu}, A^{\nu}] = i \theta^{\mu\nu}$~\cite{Ishibashi:1996xs} and those given by the generators of the $\mathfrak{so}(1,3)$ algebra~\cite{Liao:2025yfb}.
In order to perform numerical computations that consistently incorporate contributions from all such nontrivial solutions, the matrix bootstrap approach appears to be a highly promising method.

However, the discussion so far has an important limitation. Both Monte Carlo methods and bootstrap approaches have been developed mainly for Euclidean-type matrix models, and their application to Minkowski-type matrix models is far from straightforward. Here, by ``Euclidean'' and ``Minkowski'' matrix models we mean whether the weight of the matrix integral is given by $e^{-S}$ or by $e^{iS}$:
\begin{align}
Z_{\text{Euclid}} \equiv \int d\phi\, e^{-S}, 
\qquad 
Z_{\text{Minkowski}} \equiv \int d\phi\, e^{iS}.
\end{align}
A notorious difficulty in directly evaluating Minkowski-type matrix integrals is the \textit{sign problem}: the rapidly oscillating phase factor $e^{iS}$ destroys the probabilistic interpretation of the weight, rendering efficient numerical computation effectively impossible. In recent years, however, considerable effort has been devoted to methods such as the complex Langevin approach~\cite{Parisi:1983mgm, Klauder:1983sp, Aarts:2009uq} and the generalised Lefschetz thimble method~\cite{Cristoforetti:2012su, Alexandru:2015sua}, and these developments have begun to offer some hope for the direct numerical study of Minkowski-type matrix models primarily at finite $N$~\cite{Nishimura:2019qal, Chou:2025moy, Anagnostopoulos:2022dak}.

What, then, can be said about large-$N$ Minkowski-type matrix models? As a particularly intriguing example, it has been actively discussed in recent years that a $(1+3)$-dimensional spacetime may dynamically emerge from the Minkowski-type IKKT matrix model~\cite{Kim:2011cr}. 
The emergent $(1+3)$-dimensional spacetime is likely to be non-compact, in which case the theory possesses infinitely many degrees of freedom; matrix models, on the other hand, typically have $N^2$ degrees of freedom.
Therefore, theoretically, to describe a non-compact spacetime within a matrix model, it is necessary to take the limit $N \to \infty$.
At present, it is unclear how finite-$N$ effects manifest themselves in simulations of such a non-compact spacetime, and it would therefore be ideal to perform numerical computations while taking the large-$N$ limit.

In addition, zero-dimensional matrix models based on large-$N$ reduction~\cite{Eguchi:1982nm, Parisi:1982gp, GROSS1982440} are also of central importance.
Large-$N$ reduction, also known as Eguchi-Kawai equivalence, provides a framework for constructing a zero-dimensional matrix model that is equivalent to a $D$-dimensional matrix-valued field theory.
Importantly, this equivalence can be understood either as the equivalence of loop equations or as the equivalence of planar diagrams based on ribbon graphs.
Therefore, the procedure of large-$N$ reduction can be formulated, at least formally, in a manner that does not depend on the signature of spacetime.
Consequently, a Minkowski-type large-$N$ reduced matrix model is, in principle, expected to correspond to a matrix-valued field theory in $D$-dimensional Minkowski spacetime.
Such models may provide a promising avenue for exploring real-time dynamics in quantum field theory.

While bootstrap methods might appear promising for studying such large-$N$ Minkowski-type matrix models, the situation is unfortunately not so simple. As will be explained in detail in Section~\ref{sec:Bootstrap Approximation}, the positivity constraints that play a central role in bootstrap approaches break down in Minkowski-type matrix models, making it impossible to derive meaningful inequalities.

In this work, we propose a numerical method for large-$N$ Minkowski-type matrix models.
Unlike the existing matrix bootstrap approach, the present method does not use positivity at all.
Instead, we employ the \textit{eigenvalue distribution} $\rho(\lambda)$, which is characteristic of large-$N$ matrix models, and impose self-consistency constraints between this distribution and the moments $w_n = \langle \operatorname{tr} \phi^n \rangle$.
In this formulation, the constraints amount to requiring $| w_n - w_n^{(P)} | \simeq 0$.
Since this quantity is evaluated in absolute value, the resulting numerical procedure is in principle free from the sign problem and from the breakdown of positivity.

However, in Minkowski-type matrix models the weight $e^{iS}$ is complex, and therefore it is not yet fully understood what form the eigenvalue distribution $\rho(\lambda)$ should take, or even whether such an object exists.
For this reason, in the present work we consider the simplest matrix model, namely the one-matrix model, and moreover adopt the nontrivial assumption that the \textit{resolvent} $R(z)$ has only a single cut.
We then show that, at least as long as this assumption is valid, the numerical results reproduce the perturbative results with high accuracy.

The structure of this paper is as follows. 
In Section~\ref{sec:Bootstrap Approximation}, where the method is introduced, we begin with a brief review of the Hermitian one-matrix model that will serve as a toy model in this work, and then discuss positivity, which forms the theoretical foundation of usual bootstrap approaches.
In contrast to this, we introduce the eigenvalue distribution as one of the central concepts of the present study and explain how it can be approximated by polynomials.
In Section~\ref{sec:Minkowski-type model}, we discuss how this method can be extended to Minkowski-type theories.
Section~\ref{sec:Numerical Results} is devoted to concrete numerical calculations, where we demonstrate that known results for the one-matrix model are successfully reproduced in both the Euclidean and Minkowski cases.
Finally, in Section~\ref{sec:Summary and Discussion}, we summarize the method and the numerical results, and present several perspectives on how the approach can be extended to more realistic models, such as the IKKT matrix model and the (twisted) Eguchi--Kawai model.
Some technical but important calculations and facts related to the one-matrix model are collected in the Appendices.
This includes analytical (and formal) results and discussions related to the Minkowski-type one-matrix model.

\section{Bootstrap Approximation} \label{sec:Bootstrap Approximation}


In order to assess the effectiveness of our method, it is desirable to apply it to a model for which an analytic solution in the large-$N$ limit is available. Among such matrix models, the Hermitian one-matrix model is known as the simplest example that nevertheless yields nontrivial results. Its action is given by
\begin{align} \label{eq:S of one-matrix model}
S = N \operatorname{Tr}\!\left( \frac{1}{2}\phi^{2} - \frac{g}{4}\phi^{4} \right),
\end{align}
where $\phi$ is an $N\times N$ Hermitian matrix and does not depend on any spacetime coordinate $x$. In this sense, the theory may be regarded as a zero-dimensional model. Remarkably, however, in the suitable large-$N$ limit, spacetime degrees of freedom emerge from the eigenvalues of the matrix $\phi$, and the theory becomes equivalent to two-dimensional quantum gravity, namely Liouville theory. For early studies on the relation between the one-matrix model and Liouville theory, see~\cite{David:1988hj, Knizhnik:1988ak, Distler:1988jt, Brezin:1990rb, Douglas:1989ve, Gross:1989vs}, and for a comprehensive review of the random matrix theory and gravity, see~\cite{DiFrancesco:1993cyw}.

\subsection{Matrix bootstrap with positivity constraint}

Most of the existing studies on the one-matrix model have focused on Euclidean-type models. Here, by ``Euclidean'' we mean that the weight of the matrix integral is defined by $e^{-S}$, namely,
\begin{align} \label{eq:Z_E of one-matrix model}
Z_{E} \equiv \int d\phi\, e^{-S}.
\end{align}
Since $e^{-S}$ is a nonnegative real number, the expectation value of $\operatorname{Tr}(M^{\dagger}M)$, which is also nonnegative for any matrix $M$ appearing in the theory, is guaranteed to be nonnegative as well. In the case of the one-matrix model, such a matrix $M$ can be written in the form $M = \sum_{n=0}^{\infty} a_{n}\phi^{n}$, and one obtains
\begin{align} \label{eq:positivity constraint}
\langle \operatorname{Tr}(M^{\dagger}M) \rangle_{E} \ge 0.
\end{align}
This property is referred to as \textit{positivity} in matrix models, and it has served as the fundamental inequality underlying existing matrix-bootstrap approaches based on semidefinite programming (SDP).

Furthermore, in the large-$N$ limit, \textit{large-$N$ factorization} implies that expectation values of all physical observables can be expressed in terms of single-trace operators. The corresponding loop equations then take the remarkably simple form
\begin{align} \label{eq:loop eq}
g w_{n+2} &=w_{n} - \sum_{m=0}^{n-2} w_{n-m-2} w_{m} \quad \text{for} \quad n=1,2,...
\end{align}
where
\begin{align}
w_n \equiv \langle \operatorname{tr}\,\phi^n \rangle,
\quad
\operatorname{tr}\mathcal{O} \equiv \frac{1}{N}\operatorname{Tr}\mathcal{O}.
\end{align}
Due to these equations, any moment $w_{n}$ can be reduced to a combination of $w_{1}$, $w_{2}$, and $w_{0}=1$. The combination of positivity-based inequalities and the constraints on moments imposed by the loop equations constitutes the core of conventional matrix-bootstrap methods, and by implementing these conditions via SDP one can derive a nontrivial allowed region~\cite{Anderson:2016rcw, Lin:2020mme, Kazakov:2021lel}.

One of the attractive features of such matrix-bootstrap approaches is that they allow one to take the matrix size $N$ to infinity. Although loop equations in general contain terms of order $1/N^{2}$, these subleading contributions are discarded by taking the limit $N \to \infty$, so that the above equations incorporate the large-$N$ constraint from the outset. 
Given that conventional Monte Carlo methods primarily target theories with finite $N$, the two approaches may be regarded as forming an ideal complementary pair.

However, the positivity constraint~\eqref{eq:positivity constraint} introduced above clearly relies on the fact that $e^{-S} \ge 0$. In contrast, a Minkowski-type matrix model is defined by the matrix integral
\begin{align} \label{eq:Z_M of one-matrix model}
Z_{M} \equiv \int d\phi\, e^{iS},
\end{align}
where the weight $e^{iS}$ is generically a complex number. As a result,
\begin{align}
\langle \operatorname{Tr}(M^{\dagger}M) \rangle_{M} \in \mathbb{C},
\end{align}
and the notion of positivity breaks down. Indeed, if one computes, for example, $w_{2} = \langle \operatorname{tr}\phi^{2} \rangle_{M}$ perturbatively for the action \eqref{eq:S of one-matrix model}, one finds
\begin{align}
w_{2} = i - 2g - 9i g^{2} + \cdots,
\end{align}
where the leading term is purely imaginary, the first-order correction is real, and higher-order terms are neither purely real nor purely imaginary. In this way, the matrix-bootstrap approach fails already at its first step.\footnote{In theories that explicitly include time $t$ from the outset, one can define physical states $|\psi\rangle$ in an operator formalism. By requiring $\langle\psi|\psi\rangle \ge 0$, it seems possible to derive nontrivial inequalities, even if the theory is of Minkowski type. For details, see~\cite{Han:2020bkb, Lin:2023owt, Lin:2024vvg}.}

Minkowski-type matrix models nevertheless possess many intriguing features. For instance, it has been actively discussed in recent years that a $(1+3)$-dimensional spacetime dynamically emerges from the Minkowski-type IKKT matrix model. As a more direct example, Minkowski-type large-$N$ reduced matrix models are expected to correspond to Minkowski-type field theories with infinite volume~\cite{Parisi:1982gp, GROSS1982440}, and a successful analysis of such models would greatly advance our understanding of real-time dynamics of QFTs.

In view of the limitations of Monte Carlo methods---namely, their restriction to finite $N$ and the sign problem---the central requirement we impose on a matrix-bootstrap approach is the following: it should be capable of analyzing large-$N$ Minkowski-type matrix models. 

\subsection{Eigenvalue distribution}

Our method makes essential use of the eigenvalue distribution. To introduce this concept, we begin by formulating the discussion in the context of the Euclidean-type one-matrix model. The one-matrix model~\eqref{eq:S of one-matrix model} consists of a single Hermitian matrix $\phi$, and therefore all physical quantities can be expressed in terms of its eigenvalues. Let us denote the eigenvalues, ordered in ascending order, by $\lambda_{1},\ldots,\lambda_{N}$. Namely, we consider the situation
\begin{equation}
\begin{aligned}
U \phi U^{\dagger} &= \operatorname{diag}(\lambda_{1},\ldots,\lambda_{N}), \\
\lambda_{1} &\le \cdots \le \lambda_{N}, \\
\lambda_{k} &\in \mathbb{R} \qquad (k=1,\ldots,N).
\end{aligned}
\end{equation}
For Euclidean-type matrix models, the eigenvalue distribution $\rho_{E}(\lambda)$ is defined as
\begin{align} \label{eq:eigenvalue distribution}
\rho_{E}(\lambda) \equiv \left\langle \frac{1}{N} \sum_{k=1}^{N} \delta(\lambda - \lambda_{k}) \right\rangle_{E}.
\end{align}
In the limit $N \to \infty$, the eigenvalue distribution $\rho_{E}(\lambda)$ generally becomes a continuous function. Once this quantity is defined, the moments $w_{n} = \langle \operatorname{tr}\phi^{n} \rangle$ can be written as
\begin{align} \label{eq: w_n as a moment}
w_{n} = \int_{a}^{b} d\lambda\, \lambda^{n} \rho_{E}(\lambda),
\end{align}
where we have assumed that the eigenvalues are distributed over the interval $[a,b]$ and that $\rho_{E}(\lambda)=0$ outside this region. If $\rho_{E}(\lambda)$ is regarded as a probability distribution, then $w_{n}$ is nothing but the $n$-th moment in the usual sense.

All expectation values of physical observables in the one-matrix model can be expressed in terms of the moments $w_{n}$ or their combinations.
Moreover, once the eigenvalue distribution $\rho_{E}(\lambda)$ is known, the moments $w_{n}$ can be generated freely via the integral~\eqref{eq: w_n as a moment}.
Hence, the eigenvalue distribution $\rho_{E}(\lambda)$ contains all the information we need.
There exists a study~\cite{Kovacik:2025qgj} in which the moments $w_n$ are determined by a bootstrap method and the eigenvalue distribution $\rho_E(\lambda)$ is then reconstructed from them.
By contrast, our proposed method proceeds in the opposite direction: we directly determine the eigenvalue distribution $\rho_E(\lambda)$ itself by a \textit{bootstrap approximation} and subsequently compute the moments $w_n$ from it.

\subsection{Polynomial approximation of the eigenvalue distribution}

\subsubsection{Approximate ansatz}

The eigenvalue distribution $\rho_{E}(\lambda)$ is, in general, a nonnegative real function with support on the real axis (typically a finite interval), normalized so that its integral over the real axis is equal to one. For the Euclidean-type one-matrix model, the exact form of the eigenvalue distribution is known. In particular, for $g<1/12$, it is given by
\begin{align}
\rho_{E}(\lambda) & =\frac{\bigl(1-\frac{a^{2}g}{2}-g\lambda^{2}\bigr)\sqrt{(a-\lambda)(a+\lambda)}}{2\pi},\\
a & =\sqrt{\frac{2\bigl(1-\sqrt{1-12g}\bigr)}{3g}},
\end{align}
as reviewed in Appendix~\ref{appendix a}. As a first step toward a numerical approximation of this function, we propose to approximate it by a polynomial. We denote by $\rho_{E}^{(P)}(\lambda)$ the approximation of $\rho_{E}(\lambda)$ by a polynomial of degree $M$, and define the corresponding ``approximate moments'' $w_{n}^{(P)}$ as
\begin{align}
\rho_{E}^{(P)}(\lambda) &\equiv \sum_{m=0}^{M} c_{m}\lambda^{m}, \\
w_{n}^{(P)} &\equiv \int_{a}^{b} d\lambda\, \lambda^{n} \rho_{E}^{(P)}(\lambda)
= \sum_{m=0}^{M} \frac{c_{m}}{n+m+1}\left(b^{n+m+1}-a^{n+m+1}\right).
\end{align}
Here, the coefficients $c_{m}$ and the integration interval $[a,b]$ are treated as unknown variables to be determined numerically in the course of the approximation.

Let us comment on the degrees of freedom involved in this polynomial approximation. Since the exact solution above contains a square-root factor, an exact equality $\rho_{E}^{(P)}(\lambda)=\rho_{E}(\lambda)$ would require taking the limit $M\to\infty$, which would render numerical computation impossible due to the resulting infinite number of degrees of freedom. On the other hand, when a ``well-behaved'' continuous function such as~$\rho_{E}(\lambda)$ is defined on a finite interval of the real axis, it is generally well approximated by a polynomial of finite degree. From a mathematical viewpoint, this procedure amounts to regularizing the infinite degrees of freedom of a continuous function by a finite-degree polynomial.

Whenever one approximates a function, it is necessary to specify a criterion or a notion of error that defines the approximation. In case of the eigenvalue distribution $\rho_{E}(\lambda)$, we can make use of the moments $w_{n}$  and $w_{n}^{(P)}$. Precisely, the strategy is to determine the unknown variables such that the differences $|w_{n}-w_{n}^{(P)}|$ become as small as possible for all $n$.

At this stage, however, this approximation criterion remains purely formal. Since the eigenvalue distribution $\rho_{E}(\lambda)$ is unknown, its moments $w_{n}$ are also unknown, and therefore the quantities $|w_{n}-w_{n}^{(P)}|$ cannot be evaluated numerically. This obstacle can be overcome by invoking the loop equations~\eqref{eq:loop eq}. As discussed earlier, the loop equations imply that the moments $w_{n}$ can be reduced to combinations of $w_{1}$, $w_{2}$, and $w_{0}=1$. This observation leads to a key idea: if we include the moments $w_{1}$ and $w_{2}$ themselves among the unknown variables, then the differences $|w_{n}-w_{n}^{(P)}|$ become computable for all $n$. Indeed, the total number of unknown variables is
\[
\text{unknowns:}\quad c_{0},c_{1},\ldots,c_{M},a,b,w_{1},w_{2},
\]
namely $M+5$ in total, whereas the conditions $w_{n}\approx w_{n}^{(P)}$ are required to hold for all $n$, yielding infinitely many equations. The approximation problem is therefore reduced to a typical overdetermined optimization problem, which can be efficiently solved using the least-squares method. Specifically, we define the objective function
\begin{align} \label{eq:objective function}
F_E(c_{m},a,b,w_{1},w_{2}) \equiv \sum_{n=0}^{\Lambda} r_{n}\, |w_{n}-w_{n}^{(P)}|^{2},
\end{align}
and minimize it with respect to the unknown variables. Here, the coefficients $r_{n}$ are positive real numbers introduced for numerical stability. Since it is impossible to handle infinitely many equations in practice, the sum over $n$ is truncated at an upper bound $\Lambda$. In order to ensure the validity of the computation, the number of equations should not be smaller than the number of unknowns, and we therefore require $\Lambda \ge M+4$.

\subsubsection{Theoretical basis of the formulation}

It is important to emphasize that this formulation relies only on the following three assumptions: (i) there exists an eigenvalue distribution $\rho_{E}(\lambda)$ that serves as a probability distribution generating the moments $w_{n}$~(cf.~\eqref{eq: w_n as a moment}); (ii) $\rho_{E}(\lambda)$ is a continuous function with finite support and can be well approximated by a finite-degree polynomial; and (iii) the moments $w_{n}$ satisfy the loop equations~\eqref{eq:loop eq}.
From this perspective, although the method formally takes the form of a polynomial approximation to $\rho_{E}(\lambda)$, it in fact does not make use of any detailed information about $\rho_{E}(\lambda)$ itself.
Rather, by imposing the conditions (i)--(iii), one should interpret the procedure as determining $\rho_{E}(\lambda)$ and its polynomial approximation $\rho_{E}^{(P)}(\lambda)$ simultaneously.
The treatment of $w_{1}$ and $w_{2}$ as unknown variables provides clear evidence of this interpretation, since these moments are generated by $\rho_{E}(\lambda)$ and treating them as variables is effectively equivalent to determining the distribution itself.
In this sense, our method belongs to the class of bootstrap approximations, which is based on self-consistency conditions.\footnote{
A representative example is the mean-field approximation, which typically involves an ad hoc replacement of unknown variables by a mean field and may therefore fail to capture the correct physics. 
In contrast, the present method relies only on the assumptions (i)--(iii), and its theoretical foundation is therefore comparatively robust.
}

A feature that is specific to our approach concerns the existence of the eigenvalue density \(\rho_{E}(\lambda)\).
In particular, this feature plays a role in reducing the non-uniqueness of solutions to the loop equations.
In what follows, we provide an explanation of this point.

From a purely algebraic point of view, the loop equations~\eqref{eq:loop eq} form a system of nonlinear algebraic equations.
Importantly, the loop equations themselves can admit multiple solutions simultaneously.
Concretely, the loop equations alone do not determine the values of \(w_{0}, w_{1},\) and \(w_{2}\):
for arbitrary choices of these quantities, one can always construct a set of moments \(w_{n}\)
that is algebraically consistent with~\eqref{eq:loop eq}.
However, the moments \(w_{n}\) are defined through the matrix integral $w_n \equiv \int d\phi\, \operatorname{Tr}(\phi^{n})\, e^{-S}/Z$ and such an intrinsic ambiguity is not acceptable from a physical point of view.
In other words, the indeterminacy in \(w_{0}, w_{1},\) and \(w_{2}\) is an artifact of treating the loop equations~\eqref{eq:loop eq} merely as an abstract algebraic system.
To eliminate this artifact, it is necessary to take into account appropriate physical constraints.

In conventional matrix-bootstrap approaches, these physical constraints are precisely provided by the positivity constraint~\eqref{eq:positivity constraint}.
This condition implicitly incorporates the fact that the moments \(w_{n}\) are defined by a matrix integral, and it therefore makes it possible to exclude unphysical values of \(w_{0}, w_{1},\) and \(w_{2}\).
Once this is achieved, the intrinsic power of the loop equations themselves comes into play:
even seemingly trivial positivity constraints can effectively determine the values of all moments \(w_{n}\).

By contrast, in our approach the role played by positivity is replaced by the existence of the eigenvalue distribution
\(\rho_{E}(\lambda)\)\footnote{More precisely, our method assumes \(w_{0}=1\) from the outset, which should also be regarded as part of the physical constraints.}.
That is, the assumption that the moments \(w_{n}\) are generated from \(\rho_{E}(\lambda)\),
namely assumption~(i), itself functions as a physical constraint.
It is precisely this assumption that allows us to include the moments \(w_{1}\) and \(w_{2}\) in the objective function \(F\),
and to extract their (approximate) values through a least-squares procedure.

In this paper, we do not pursue the issue of non-uniqueness any further.
Instead, because our procedure is explicitly based on the existence of the eigenvalue density \(\rho_{E}(\lambda)\), we regard the resulting approximate values as strong candidates for physical solutions, at least in the Euclidean-type one-matrix model.
At the same time, it should be kept in mind that, independently of the issue of non-uniqueness,
the present method is intrinsically approximate.
In order to justify the resulting approximate values as candidates for physical solutions, it is first necessary to verify that this approximation indeed provides a (algebraic) solution to the loop equations, as will be discussed in Section~\ref{subsec:Consistency Check of the Approximate Results}.

\subsection{Consistency check of the approximate results} \label{subsec:Consistency Check of the Approximate Results}

Since the Euclidean one-matrix model is exactly solvable, the validity of an approximation can be assessed by directly comparing the approximate eigenvalue distribution and moments with the exact solution. However, for most other matrix models, such exact solutions are not known. We therefore propose two alternative consistency checks.

The first consistency check is based on the loop equations. 
In this study, we choose to balance the number of equations and the number of unknown variables.
As a consequence, it is nontrivial whether the eigenvalue distribution $\rho_{E}^{(P)}(\lambda)$ determined in this way, as well as the moments $w_{n}^{(P)}$ generated from it, satisfy the loop equations in the region beyond the cutoff $\Lambda$.
More precisely, the unknown variables $c_{m},a,b,w_{1},w_{2}$ are trained so as to minimize $|w_{n}-w_{n}^{(P)}|^{2}$ for $n=0,\dots,\Lambda$, whereas the values of $|w_{n}-w_{n}^{(P)}|^{2}$ for $n>\Lambda$ are not used in the training.
One might therefore naively expect that the function
\begin{equation}
G'(c_{m},a,b,w_{1},w_{2})\equiv\sum_{n=\Lambda+1}^{\Lambda'}|w_{n}-w_{n}^{(P)}|^{2}
\end{equation}
could be used as a test.
However, this choice has a drawback.
In general, both $w_{n}$ and $w_{n}^{(P)}$ tend to grow (often extremely rapidly) as $n$ increases.
This is because they are generated by integrals of the form $\int_{-a}^{a}d\lambda\,\lambda^{n}\rho(\lambda)$, and for large $n$ they are expected to scale roughly as $a^{n}$.
As a result, for large $n$, the quantity $|w_{n}-w_{n}^{(P)}|^{2}$ naturally becomes large simply due to the magnitude of $w_{n}$ and $w_{n}^{(P)}$ themselves.
In such a situation, the test would be dominated by large-$n$ contributions, thereby reducing its effectiveness.
To avoid this problem, it is natural to normalize $|w_{n}-w_{n}^{(P)}|^{2}$ by $|w_{n}|^{2}$ before summing:
\begin{equation} \label{eq:test function G}
G(c_{m},a,b,w_{1},w_{2})\equiv\sum_{n=\Lambda+1}^{\Lambda'}\frac{|w_{n}-w_{n}^{(P)}|^{2}}{|w_{n}|^{2}}.
\end{equation}
In the present check, we introduce a threshold $G\le 1$ for the test function $G$, and use $L\equiv \Lambda'-\Lambda$ for the largest $\Lambda'$ satisfying this inequality as an indicator of the quality of the approximation.
A larger $L$ implies that higher-order contributions $|w_{n}-w_{n}^{(P)}|^{2}$ are more strongly suppressed, and hence that the approximation performs better.

Note that we assess the quality of the approximation using $L$, rather than the value of $G$ itself.
This is because, in the present one-matrix model, the discrepancy between $w_{n}$ and $w_{n}^{(P)}$ grows rapidly as $n$ increases, and beyond a certain point the ratio $\frac{|w_{n}-w_{n}^{(P)}|^{2}}{|w_{n}|^{2}}$ becomes almost unity.

The second approach makes use of the positivity constraint.
Since the present approximation scheme does not employ positivity, it is a priori unclear whether the resulting approximate solution satisfies the positivity constraint, and this can therefore serve as a nontrivial consistency check.
To formulate this check, following the approach of~\cite{Lin:2020mme, Kazakov:2021lel}, we first define the infinite-dimensional Hankel matrix $H$ as follows:
\begin{align}
H_{nm} \equiv w_{n+m}.
\end{align}
Using this definition, the positivity constraint can be rewritten as
\begin{equation}
\langle \operatorname{Tr} M^{\dagger} M \rangle
= \sum_{n,m} c_n^{*} c_m w_{n+m}
= \vec{c}^{\dagger} H \vec{c}
\ge 0
\quad \text{for all } \vec{c}.
\end{equation}
This condition is equivalent to
\begin{equation}
H \succcurlyeq 0,
\end{equation}
namely, that $H$ is positive semidefinite (all eigenvalues are non-negative).
The matrix $H$ itself is infinite dimensional and thus not directly suitable for numerical analysis.
However, by a simple manipulation, one can extract an arbitrary number of meaningful inequalities.

Concretely, the positivity constraint requires that $\vec{c}^{\dagger} H \vec{c}$ be non-negative for all infinite-dimensional complex vectors $\vec{c}$.
If we restrict $\vec{c}$ to vectors of the form $\vec{c}_{(k)} = (c_0, \ldots, c_{k-1}, 0, 0, \ldots)$, this expression reduces to $\vec{c}_{(k)}^{\dagger} H^{(k)} \vec{c}_{(k)}$, where $H^{(k)}$ denotes the $k$-dimensional principal submatrix of $H$.
Since positivity requires $\vec{c}^{\dagger} H \vec{c} \ge 0$ for all $\vec{c}$, it follows that $\vec{c}_{(k)}^{\dagger} H^{(k)} \vec{c}_{(k)} \ge 0$ must hold for all $\vec{c}_{(k)}$.
In other words, the latter is a necessary condition for the former, and we obtain
\begin{equation}
H \succcurlyeq 0
\ \Rightarrow \
\forall k, \quad H^{(k)} \succcurlyeq 0 .
\end{equation}
Unlike the original infinite-dimensional matrix $H$, each $H^{(k)}$ is finite dimensional and can therefore be handled numerically.
One can then combine this condition with the loop equations. 
More explicitly, for
\begin{align}
H^{(k)}(w_n)
\equiv
\begin{pmatrix}
w_0 & w_1 & \cdots & w_{k-1} \\
w_1 & w_2 &        &        \\
\vdots &     & \ddots &     \\
w_{k-1} &     &        & w_{2(k-1)}
\end{pmatrix},
\end{align}
we apply the loop equations~\eqref{eq:loop eq} to express all matrix elements in terms of $w_1$ and $w_2$.
Imposing the condition $H^{(k)} \succcurlyeq 0$, namely that all principal minors of $H^{(k)}$ be non-negative, then yields inequalities involving $w_1$ and $w_2$.
In principle, ignoring computational cost, one may take $k$ to be any natural number and derive an arbitrarily large set of inequalities for $w_1$ and $w_2$.
By combining all these inequalities using semidefinite programming (SDP), one can determine the allowed region in the $(w_1, w_2)$ plane.

The entire procedure described above relies crucially on the existence of the positivity constraint, whereas our method does not use it at all.
Therefore, the condition $H^{(k)} \succcurlyeq 0$ can instead be employed as a consistency check of the approximation.
A naive implementation proceeds as follows.
First, approximate values $w_n^{(P)}$ are obtained by the least-squares method.
These values are then substituted for all $w_n$ appearing in $H^{(k)}$, and one checks whether the resulting matrix satisfies the positivity constraint, that is,
\begin{align}
H^{(k)}(w_n^{(P)})
\equiv
\begin{pmatrix}
w_0^{(P)} & w_1^{(P)} & \cdots & w_{k-1}^{(P)} \\
w_1^{(P)} & w_2^{(P)} &        &               \\
\vdots    &           & \ddots &               \\
w_{k-1}^{(P)} &        &        & w_{2(k-1)}^{(P)}
\end{pmatrix}
\end{align}
satisfies
\begin{equation}
\forall k\in\{1,...,K\},\quad \det \!\left[ H^{(k)}(w_n^{(P)}) \right] \ge 0.
\end{equation}

Unfortunately, for the one-matrix model, this check does not turn out to be very restrictive.
The reason can be explained as follows.
As is already shown, the condition $\vec{c}_{(k)}^{\dagger} H^{(k)} \vec{c}_{(k)} \ge 0$ is a necessary condition for positivity.
Since we assume that the moments $w_n$ are generated by an eigenvalue distribution $\rho_E(\lambda)$, substituting its representation yields
\begin{equation}
\vec{c}_{(k)}^{\dagger} H^{(k)} \vec{c}_{(k)}
=
\int_{a}^{b} d\lambda
\left| \sum_{n=0}^{k-1} c_n \lambda^{n} \right|^{2}
\rho_E(\lambda).
\end{equation}
Here $[a,b]$ denotes the support of $\rho_E(\lambda)$, which lies on the real axis.
Since the factor $\left| \sum_{n=0}^{k-1} c_n \lambda^{n} \right|^{2}$ is non-negative for all $\lambda$, it follows that if $\rho_E(\lambda)$ is non-negative on its support, then the above expression is non-negative for any $\vec{c}_{(k)}$, and hence for any $\vec{c}$.
Thus, assuming the existence of the eigenvalue distribution $\rho_E(\lambda)$, the positivity constraint can be rewritten simply as
\begin{equation}
\forall \lambda \in [a,b], 
\quad \rho_E(\lambda) \ge 0.
\end{equation}
The same argument applies if one replaces $H^{(k)}$ by $H^{(k)}(w_n^{(P)})$.
Therefore, as long as the approximate eigenvalue distribution $\rho_E^{(P)}(\lambda)$ is non-negative on its support, the condition $H^{(k)}(w_n^{(P)}) \succcurlyeq 0$ is automatically satisfied for any $k$.

To make the consistency check more powerful, one can use the loop equations.
The partial Hankel matrix $H^{(k)}(w_{n})$ contains $w_{0},...,w_{2(k-1)}$ as its entries, but as mentioned above, all these moments can be reduced to combinations of $w_{1}$ and $w_{2}$ by the loop equations.
We distinguish this representation by writing $\bar{H}^{(k)}(w_{1},w_{2})$.
To be precise,
\begin{equation}
\bar{H}^{(k)}(w_{1},w_{2})\overset{\star}{=}H^{(k)}(w_{n})
\end{equation}
is the definition of $\bar{H}^{(k)}(w_{1},w_{2})$.
Here ``$\overset{\star}{=}$'' indicates that all $w_{n}$ are reduced to $w_{1}$ and $w_{2}$ using the loop equations.
One then replaces $w_{1},w_{2}$ by the approximate values $w_{1}^{(P)},w_{2}^{(P)}$, and checks whether
\begin{equation}
\forall k\in\{1,...,K\},\ \ \ \det[\bar{H}^{(k)}(w_{1}^{(P)},w_{2}^{(P)})]\ge0
\end{equation}
holds.
One then looks for the maximal $K$ for which this condition is satisfied.

This check is almost a direct transplantation of the method used in the standard matrix bootstrap into a consistency check, and it is expected to serve as a reliable indicator for judging the success or failure of the approximation.
However, as discussed earlier, the positivity constraint can be properly defined only for Euclidean matrix models at present. It should therefore be noted that this consistency check cannot be applied to Minkowski-type matrix models. Instead, in this work we compare the approximation results with those obtained from perturbative calculations; this comparison is important for determining whether the resulting approximations merely correspond to algebraic solutions of the loop equations.

\section{Extension to the Minkowski-type models} \label{sec:Minkowski-type model}

The least-squares formulation is designed so that the absolute difference between the two moments, $w_{n}$ and $w_{n}^{(P)}$, is driven to zero. As a result, neither the sign problem inherent in Monte Carlo methods nor the breakdown of positivity in usual bootstrap approaches arises at a fundamental level. Nevertheless, in order to apply this method to Minkowski-type models, it is necessary to further clarify the theoretical foundations.

Let us recall the definitions of the eigenvalue distribution $\rho_{E}(\lambda)$ in the Euclidean case~\eqref{eq:eigenvalue distribution}. This is defined as the expectation value of a sum of delta functions and has finite support $[a,b]$ on the real axis. How to extend this notion to Minkowski-type theories is a nontrivial issue. The reason why $\rho_{E}(\lambda)$ has support on the real axis is that the matrix variable $\phi$, as well as its expectation value $\langle \phi \rangle_{E}$ weighted by $e^{-S}$, is Hermitian, ensuring the reality of its eigenvalues. In contrast, the Minkowski-type expectation value $\langle \phi \rangle_{M}$, weighted by $e^{iS}$, is no longer Hermitian, and its eigenvalues may therefore become complex.

This implies that, if one attempts to define an eigenvalue distribution $\rho_{M}(\lambda)$ for the Minkowski-type one-matrix model, it can be a complex-valued function defined on a curve $\Gamma$ in the complex plane. A related technical difficulty is that a one-dimensional delta function $\delta(z)$ on the complex plane is, in general, not well defined. Typically, delta functions on the complex plane must be introduced as distributions with respect to an area integral, namely $\delta^{2}(z)=\delta(\operatorname{Re}z)\delta(\operatorname{Im}z)$.

Summarizing these points, one would like to extend the Euclidean definitions to the Minkowski case in the form
\begin{align}
\rho_{M}(z) &\overset{?}{=} \left\langle \frac{1}{N}\sum_{k=1}^{N} \delta(z-\lambda_{k}) \right\rangle, \\
w_{n} &\overset{?}{=} \int_{\Gamma} dz\, z^{n}\rho_{M}(z),
\end{align}
but the problem is that it is unclear whether these statements are actually correct, or even whether they are well-defined from a mathematical point of view.

\subsection{Large-\texorpdfstring{$N$}{N} master field: a ``density-matrix'' interpretation}
\label{subsubsec:density-matrix interpretation}

To apply the bootstrap approximation method to Minkowski-type theories,
we first revisit and reinterpret the eigenvalue distribution in the Euclidean theory.
For this purpose, we begin by introducing the concept of the \textit{large-$N$ master field}.
In matrix models, the relation
\begin{align}
\langle O_1 O_2\rangle - \langle O_1\rangle \langle O_2\rangle
\sim \mathcal O(N^{-2}),
\end{align}
namely the \textit{large-$N$ factorization} property,
holds in a wide class of models
for quantities $O_1$ and $O_2$
written in terms of the normalized trace
$\text{tr}(\cdots) = \frac{1}{N}\text{Tr}(\cdots)$.
From this relation, it follows, for example, that for any matrix $A_\mu$ appearing in the theory,
\begin{align}
\forall n\in\mathbb N,\quad
\langle (\text{tr}\, A_\mu)^n\rangle
- \langle \text{tr}\, A_\mu\rangle^n
\sim \mathcal O(N^{-2}).
\end{align}
Taking the large-$N$ (planar) limit, the right-hand side vanishes.
This implies that if $\text{tr}\, A_\mu$ is regarded as a random variable,
its fluctuations completely disappear in the $N\to\infty$ limit.
Moreover, the same statement holds not only for $\text{tr}\, A_\mu$
but also for any normalized single-trace operator $\text{tr}(A_{\mu_1}\cdots A_{\mu_m})$.
From this simple observation, one is led to the expectation that
the matrix integral is dominated by a single saddle point in the large-$N$ limit. This expectation suggests the existence of matrices $\tilde A_\mu$,
namely a large-$N$ master field, satisfying
\begin{align} \label{eq:master field 1}
\langle \text{tr}(A_{\mu_1}\cdots A_{\mu_m})\rangle
\overset{N\to\infty}{=}
\text{tr}(\tilde A_{\mu_1}\cdots \tilde A_{\mu_m}).
\end{align}
More precisely, the large-$N$ master-field conjecture states that,
for the matrices $A_\mu$ in the theory,
there exist corresponding operators $\tilde A_\mu$ in the $N\to\infty$ limit
such that the expectation value of any single-trace operator
$\text{tr}(A_{\mu_1}\cdots A_{\mu_m})$
is equal to the trace
$\text{tr}(\tilde A_{\mu_1}\cdots \tilde A_{\mu_m})$.
In other words, although expectation values are statistical averages taken with respect to the weights $e^{-S}$ or $e^{iS}$, they behave deterministically, or ``classically,''
in the large-$N$ limit.

While~\eqref{eq:master field 1} is intuitively clear, in the $N\to\infty$ limit
the factor $1/N$ vanishes, and hence,
as long as the left-hand side remains finite,
$\text{Tr}(\tilde A_{\mu_1}\cdots \tilde A_{\mu_m})$
diverges proportionally to $N$.
In this situation, it is a nontrivial problem to explicitly construct
the individual matrices $\tilde A_\mu$,
and such singular large-$N$ behavior is also unsuitable for numerical computations.
We therefore rewrite the master-field relation in the following form:
\begin{align} \label{eq:master field 2}
\langle \operatorname{tr}(A_{\mu_1}\cdots A_{\mu_m})\rangle
\overset{N\to\infty}{=}
\langle\Omega|\hat A_{\mu_1}\cdots \hat A_{\mu_m}|\Omega\rangle_{\cal H}
=
\operatorname{Tr}_{\cal H}\!\left(\hat A_{\mu_1}\cdots \hat A_{\mu_m}\hat\rho\right).
\end{align}
Here ${\cal H}$ is a representation space, typically infinite-dimensional, and $\operatorname{Tr}_{\cal H}$ is not assumed to arise as the $N\to\infty$ limit of the original $N$-dimensional trace.
In the Euclidean case, if the planar limits of all single-trace moments exist, they define a positive linear functional (a state) on the $\ast$-algebra generated by the observables.
The existence of such a Hilbert space ${\cal H}$, a vector $|\Omega\rangle\in{\cal H}$, and operators $\hat A_\mu$ realizing the planar state is then guaranteed by the GNS construction.

In the last expression we take $\hat\rho:=|\Omega\rangle\langle\Omega|$, which formally resembles a density matrix.
In the following we therefore refer to $\hat\rho$ simply as a ``density matrix''.
We emphasize, however, that~\eqref{eq:master field 2} is merely an alternative representation of~\eqref{eq:master field 1}, and $\hat\rho$ is not required to represent a physical density matrix.
We will not pursue any further physical interpretation of the master field $\hat A_\mu$ or the density matrix $\hat\rho$, and use them solely as a convenient mathematical device for implementing the bootstrap approximation.

This form of the master field and density matrix in~\eqref{eq:master field 2} can be explicitly constructed
in the Euclidean one-matrix model.
In the one-matrix model, the expectation values
$\langle \text{tr}(A_{\mu_1}\cdots A_{\mu_m})\rangle$
correspond to the moments
$w_n=\langle \text{tr}\,\phi^n\rangle$,
and hence the operators $\hat\phi$ and $\hat\rho_E$ are required to satisfy
\begin{align} \label{eq:master field conjecture of 1-m.m.}
w_n=\text{Tr}_{\mathcal H}(\hat\phi^n\hat\rho_E),
\quad n=0,1,2,\dots .
\end{align}
This relation closely resembles the relation between $w_n$
and the eigenvalue density $\rho_E(\lambda)$ in~\eqref{eq: w_n as a moment},
and indeed one can make use of this relation to construct $\hat\phi$ and $\hat\rho_E$.
First let us introduce a complete orthonormal basis
$\{|\lambda\rangle\}$ of the Hilbert space $\mathcal H$
satisfying
\begin{equation}  \label{eq:complete set}
\begin{aligned}
\langle\lambda|\lambda'\rangle &= \delta(\lambda-\lambda'),\\
1 &= \int_{-\infty}^{\infty} d\lambda\,|\lambda\rangle\langle\lambda|.
\end{aligned}
\end{equation}
We then define $\hat\phi$ and $\hat\rho_E$ so that
the following eigenvalue equations hold:
\begin{equation} \label{eq:master field and density matrix}
\begin{aligned} 
\hat\phi|\lambda\rangle &= \lambda|\lambda\rangle,\\
\hat\rho_E|\lambda\rangle &= \rho_E(\lambda)|\lambda\rangle.
\end{aligned}
\end{equation}
Given the completeness of $\{|\lambda\rangle\}$,
the trace can be evaluated as
$\text{Tr}_{\mathcal H}A
= \int_{-\infty}^{\infty} d\lambda\,\langle\lambda|A|\lambda\rangle$.
By inserting the resolution of the identity
$\int_{-\infty}^{\infty} d\lambda_k\,|\lambda_k\rangle\langle\lambda_k|=1$
$n$ times, one immediately finds
\begin{equation} 
\begin{aligned}
\text{Tr}_{\mathcal H}(\hat\phi^n\hat\rho_E)
&=
\int_{-\infty}^{\infty}
d\lambda\, d\lambda_1\cdots d\lambda_n\,
\langle\lambda|\hat\phi|\lambda_1\rangle
\cdots
\langle\lambda_n|\hat\rho_E|\lambda\rangle\\
&=
\int_a^b d\lambda\,\lambda^n\rho_E(\lambda)
= w_n .
\end{aligned}
\end{equation}
Thus, $\hat\phi$ indeed serves as the master field, and $\hat\rho_E$ as the density matrix.
Equivalently, one may define
\footnote{
To reproduce the vector $|\Omega\rangle$ appearing in~\eqref{eq:master field 2}, one may choose
$|\Omega\rangle=\int d\lambda\,\sqrt{\rho_E(\lambda)}\,|\lambda\rangle$.
The corresponding rank-one operator is then given by
$|\Omega\rangle\langle\Omega|
=\int d\lambda\,d\lambda'\,\sqrt{\rho_E(\lambda)\rho_E(\lambda')}\,
|\lambda\rangle\langle\lambda'|$,
which contains off-diagonal components in the $\{|\lambda\rangle\}$ basis.
However, since $\hat\phi$ is diagonal in this basis, only the diagonal components contribute to expectation values of the form
$\langle\Omega|\hat\phi^n|\Omega\rangle$.
}
\begin{equation}
\begin{aligned} 
\hat\phi &\equiv \int_{-\infty}^{\infty} d\lambda\,
\lambda\,|\lambda\rangle\langle\lambda|,\\
\hat\rho_E &\equiv \int_{-\infty}^{\infty} d\lambda\,
\rho_E(\lambda)\,|\lambda\rangle\langle\lambda|.
\end{aligned}
\end{equation}

In the Euclidean one-matrix model, the vectors $|\lambda\rangle$
are labeled by the eigenvalues $\lambda$,
the master field $\hat\phi$ has $|\lambda\rangle$ as eigenvectors
with eigenvalues $\lambda$,
and the density matrix is given by a superposition of
$|\lambda\rangle\langle\lambda|$
weighted by $\rho_E(\lambda)$.
However, when this construction is extended to multi-matrix models,
it is far from obvious whether $\hat A_\mu$ and $\hat\rho$
admit such a transparent interpretation.
We emphasize again that our standpoint is to use this master-field representation
or density-matrix interpretation solely as a tool for the bootstrap approximation,
and we do not pursue its physical interpretation further in this paper.

We now turn to how this density-matrix interpretation can be exploited
for numerical computations in Minkowski-type matrix models
(the extension to multi-matrix models is discussed in Section~\ref{sec:Summary and Discussion}).
In the discussion above, we constructed the density matrix $\hat\rho_E$
starting from the eigenvalue density $\rho_E(\lambda)$.
However, under the master-field conjecture~\eqref{eq:master field 2},
the more fundamental object is $\hat\rho_E$ itself,
while $\rho_E(\lambda)$ emerges as its eigenvalue distribution.
Motivated by the existence of the large-$N$ factorization, we therefore assume that an analogous relation holds also in the Minkowski one-matrix model, namely
\begin{align} \label{eq:master field for m-type 1-m.m.}
\text{Tr}_{\mathcal H}(\hat\phi^n\hat\rho_M)=w_n .
\end{align}
It should be noted that $\hat\rho_M$ and $\hat\phi$
are operators weighted by $e^{iS}$,
and hence they are not necessarily Hermitian
in the Minkowski theory.
Following the construction in the Euclidean case~\eqref{eq:master field and density matrix},
we assume that an analogous representation also exists
in the Minkowski theory:
\begin{equation} \label{eq:ansatz for m-type 1-m.m.}
\begin{aligned}
\hat\phi|\lambda\rangle &= \phi(\lambda)|\lambda\rangle,\\
\hat\rho_M|\lambda\rangle &= \bar \rho_M(\lambda)|\lambda\rangle.
\end{aligned}
\end{equation}
Here $\bar \rho_M(\lambda)$ is a complex-valued function
defined only on the real interval $[a,b]$.
The restriction to the real axis reflects the fact that
we take $\lambda\in\mathbb R$ (cf.~\eqref{eq:complete set}).

The crucial difference from the Euclidean case lies in $\phi(\lambda)$.
In the Euclidean theory, the relation
$w_n=\int d\lambda\,\lambda^n\rho_E(\lambda)$
is known from the outset,
so one can simply set
$\hat\phi|\lambda\rangle=\lambda|\lambda\rangle$ so as to satisfy~\eqref{eq:master field conjecture of 1-m.m.}.
In the Minkowski case, however,
no such expression for $w_n$ is known,
and it is therefore safer to treat $\phi(\lambda)$
as a general function of $\lambda$.
One may, for example, approximate it by a polynomial
and determine it numerically by a least-squares method.
As explained in Appendix~\ref{appendix b}, however,
in the present case we are led to the expectation that
\begin{align} \label{eq:phi is phase}
\phi(\lambda)=e^{i\theta}\lambda
\end{align}
holds.
Here $\theta$ is an unknown real constant to be determined by the bootstrap approximation.
In other words, we adopt the \textit{single-cut ansatz} as a working hypothesis.
This choice is motivated by the fact that it reproduces the perturbative expansion around $g=0$, and it is further assessed \emph{a posteriori} by the consistency measures $F_M$ and $G_M$ and by comparison with the formal solution. 
In what follows, we primarily proceed with the analysis under the assumption $\phi(\lambda) = e^{i\theta}\lambda$.  

Using the completeness of $\{|\lambda\rangle\}$,
the trace $\text{Tr}_{\mathcal H}(\hat\phi^n\hat\rho_M)$
can be converted into an integral as
\begin{align}
\text{Tr}_{\mathcal H}(\hat\phi^n \hat\rho_M)
=\int_a^b d\lambda\,(e^{i\theta}\lambda)^n\bar\rho_M(\lambda).
\end{align}
Introducing a new variable $z$ and a new function $\bar\rho_M(z)$ by
\begin{equation}
\begin{aligned}
z &= e^{i\theta}\lambda,\\
\rho_M(z) &= e^{-i\theta}\bar\rho_M(e^{-i\theta}z),
\end{aligned}
\end{equation}
this expression can be rewritten as
\begin{align} \label{eq:w_n in m-type 1-m.m. given by master field}
\text{Tr}_{\mathcal H}(\hat\phi^n \hat\rho_M)
=\int_\Gamma dz\,z^n\rho_M(z),
\end{align}
where $\Gamma$ is a line segment in the complex plane passing through the origin, with endpoints $e^{i\theta}a$ and $e^{i\theta}b$, and forming an angle $\theta$ with the real axis.


Finally, it should be noted that the discussion so far involves several important and nontrivial assumptions.
Namely, (1) that the master field $\hat{\phi}$ exists, (2) that, as in the Euclidean case, the eigenvectors $|\lambda\rangle$ of $\hat{\phi}$ form a complete basis and at the same time are also eigenvectors of $\hat{\rho}_{M}$, and (3) that the eigenvalues of $\hat{\phi}$ are given by $e^{i\theta}\lambda$.
Among these, the third condition corresponds to the assumption that the resolvent $R(z)$ of the Minkowski-type one-matrix model has a single cut in the complex plane.
Under this assumption the perturbative expansion of the moment $w_{2}$ can be reproduced exactly, and therefore the assumption is expected to be valid at least in the region where the absolute value of the coupling constant $g$ is small (see the Appendix~\ref{appendix b} for the detailed calculation).

However, in the region where $g$ becomes large, perturbation theory can no longer be trusted in a naive way, and the assumption of the one-cut structure loses its justification.
This implies that, once the above three assumptions are imposed, the numerical computation necessarily loses a certain degree of generality.
In particular, for relatively large values of $g$, where the perturbative expansion around $g=0$ is no longer expected to provide a reliable guide, even if the optimization converges successfully, it does not necessarily follow that the obtained solution corresponds to a genuinely physical one.
This issue must ultimately be examined by a method that does not rely on the one-cut assumption, and a detailed investigation will be presented in a forthcoming paper.

\subsection{Implementation of the Numerical Computation}

Under the least-squares method, $\bar{\rho}_{M}(z)$ is approximated by a finite-degree polynomial, in the same manner as in the Euclidean case.
Currently the moments $w_{n}$ in the Minkowski-type one-matrix model are supposed to be given by the right-hand side of~\eqref{eq:w_n in m-type 1-m.m. given by master field}. Therefore, in complete analogy with the Euclidean case, we introduce the polynomial approximation
\begin{align}
\rho_{M}^{(P)}(z) &\equiv \sum_{m=0}^{M} c_{m} z^{m}, \\
w_{n}^{(P)} &\equiv \int_{e^{i\theta}a}^{e^{i\theta}b} dz\, z^{n}\rho_{M}^{(P)}(z)
= \sum_{m=0}^{M} \frac{c_{m} e^{i\theta(n+m+1)}}{n+m+1}\left(b^{n+m+1}-a^{n+m+1}\right).
\end{align}
Here $a$, $b$, and $\theta$ are taken to be real, whereas the coefficients $c_{m}$ and the moments $w_{n}$ must in general be treated as complex numbers. The parameter $\theta$ is an additional real unknown variable that should be determined numerically as well.

Using these definitions, we introduce the objective function
\begin{align} \label{eq:objective function in Minkowski}
F_{M}(c_{m},a,b,\theta,w_{1},w_{2}) \equiv \sum_{n=0}^{\Lambda} r_{n}\, |w_{n}-w_{n}^{(P)}|^{2},
\end{align}
and perform a least-squares minimization so that $F_{M}$ becomes as close to zero as possible. Note that we again need to impose that the number of equations is larger than the number of real unknowns. Counting the unknowns in terms of real degrees of freedom, we now have $2M+9$ variables in total, which is roughly twice as many as in the Euclidean case. On the other hand, each condition $w_{n}-w_{n}^{(P)}\approx 0$ provides two real equations (from its real and imaginary parts), and therefore the lower bound on $\Lambda$ is almost the same as in the Euclidean setup.
The moments $w_n$ appearing in the objective function are reduced to combinations of $w_1$ and $w_2$ by using the loop equation
\begin{align} \label{eq:loop eq in Minkowski}
0=\sum_{m=0}^{n-2} w_{n-m-2} w_m + i\bigl(w_n - g w_{n+2}\bigr).
\end{align}

\subsection{Regularization}

Finally, we briefly comment on the regularization of Minkowski-type matrix models.
Minkowski-type models are defined by matrix integrals of the form given in~\eqref{eq:Z_M of one-matrix model}, and as long as the action $S$ is real, these integrals are purely oscillatory.
This situation also appears in general quantum field theory, where such integrals are interpreted using the $i\epsilon$ prescription.
From a mathematical viewpoint, this corresponds to the Fresnel integral, and by agreeing to take the limit $\epsilon \to +0$ after performing the integration, one can assign a meaning to integrals weighted by $e^{iS}$.

However, this kind of regularization can sometimes be problematic.
For example, in the IKKT matrix model, it was argued in~\cite{Asano:2024def} that an IR regulator induces a ``classicalization'' effect, thereby altering the theory itself.
Furthermore, for the so-called Polarised IKKT matrix model~\cite{Bonelli:2002mb}, which is obtained by introducing a mass deformation into the IKKT model, it was pointed out in~\cite{Komatsu:2024ydh} that the massless limit does not coincide with the original IKKT model.
These issues may be related to the subtlety of introducing an external scale into the scale-free IKKT model and then taking that scale to zero.

Turning to the treatment of the Minkowski-type one-matrix model in the present work, no explicit regularization based on the $i\epsilon$ prescription has been carried out so far.
This is because our stance in this study is to find consistent solutions of the loop equations~\eqref{eq:loop eq} and~\eqref{eq:loop eq in Minkowski}, and to demonstrate that such solutions are consistent with the existence of a (formal) eigenvalue distribution.
Within this framework, it is not necessary to explicitly invoke the matrix integral itself.
In addition, as seen in Section~\ref{sec:Numerical Results}, the present bootstrap approximation for the Minkowski-type one-matrix model reproduces perturbative results well, and moreover appears to provide a good fit to the formal eigenvalue distribution $\rho_{M}(\lambda)$.

Nevertheless, it would be unwarranted to conclude from this that Minkowski-type matrix models generally do not require the $i\epsilon$ prescription.
The most delicate point is that, without the $i\epsilon$ prescription, matrix integrals weighted by $e^{iS}$ are in general ill-defined, and therefore the derivation of loop equations from them may not be justified.
In the case of the one-matrix model, if one first introduces the $i\epsilon$ prescription to ensure convergence of the integral, then derives the loop equations, and only afterwards takes the limit $\epsilon \to +0$, the effect of the $i\epsilon$ term seems to disappear completely.
(Here it should be noted that, unlike the IKKT matrix model, the one-matrix model possesses an intrinsic scale originating from the coupling constant $g$.)
For the time being, in this work we adopt the safe standpoint that the loop equations are obtained after performing the standard $i\epsilon$ prescription, and we leave the subtle issues related to the $i\epsilon$ regularization of Minkowski-type matrix models to future studies.

\section{Numerical Results} \label{sec:Numerical Results}

\subsection{Overall Setup} \label{sebsec:overall setup}

In this work, we performed the least-squares minimization using the \texttt{least\_squares} function provided by SciPy~\cite{virtanen2020scipy}. We employed the \texttt{trf} algorithm, and set \texttt{ftol}, \texttt{xtol}, and \texttt{gtol} all to $10^{-14}$, with \texttt{max\_nfev} set to $20000$. For simplicity and numerical stability, we assume the $\mathbb{Z}_{2}$ symmetry of the theory throughout all numerical computations. Namely, we impose
\begin{align}
w_{2n-1}=0 \qquad (n=1,2,\ldots)
\end{align}
for all odd moments. Since the loop equations~\eqref{eq:loop eq} are completely decoupled between odd and even moments, this assumption does not affect the dynamics of the even moments $w_{2n}$.

Under this assumption, the eigenvalue distribution necessarily becomes an even function, and the polynomial approximation $\rho^{(P)}(\lambda)$ and the corresponding approximate moments $w_{n}^{(P)}$ reduce to
\begin{equation} \label{eq: approximation anzats}
\begin{aligned}
\rho^{(P)}(\lambda) &\equiv \sum_{m=0}^{M/2} c_{2m}\lambda^{2m}, \\
w_{n}^{(P)} &\equiv \int_{-a}^{a} d\lambda\, \lambda^{n}\rho^{(P)}(\lambda).
\end{aligned}
\end{equation}
As a consequence, odd-order coefficients $c_{2m-1}$ need not be considered, and the integration domain can be chosen to be the symmetric interval $[-a,a]$. At the same time, all loop equations for the odd moments $w_{2n-1}$ become trivial, so that the number of equations is reduced accordingly.

Taking the $\mathbb{Z}_{2}$ symmetry into account, the independent variables of the theory are $w_{2}$, $a$, $\theta$, and the coefficients $c_{2m}$. In the Minkowski case, $w_{2}$ and $c_{2m}$ are treated as complex variables, whereas in the Euclidean case the parameter $\theta$ is unnecessary and all variables are taken to be real. In order to perform the optimization, initial conditions must be specified. For both the Euclidean and Minkowski cases, we chose
\begin{equation}
\begin{aligned}
w_{2}^{\text{initial}} &= 1, \\
a^{\text{initial}} &= 1, \\
\theta^{\text{initial}} &= 0, \\
c_{2m}^{\text{initial}} &=
\begin{cases}
1 & (m=0), \\
0 & \left(m=1,\ldots,\frac{M}{2}\right).
\end{cases}
\end{aligned}
\end{equation}
The optimization typically converges very rapidly. Even in the Minkowski case, where the number of variables is larger, the computation finishes well within one minute for \texttt{max\_nfev}$=20000$. In practice, the optimization often converges after approximately $2000$ iterations, depending on the choice of initial conditions.

For the objective function $F$ in~\eqref{eq:objective function} and~\eqref{eq:objective function in Minkowski}, an upper cutoff $\Lambda$ is imposed on the order $n$ of the moments $w_n$. In principle, one could take $\Lambda$ to be large and thereby formulate an overdetermined system. However, due to the nature of the least-squares method, this increases the risk of the optimization being trapped in local minima that do not correspond to the true minimum. We therefore choose to match the number of unknown variables with the number of equations. 

This option also has a drawback. 
When the number of unknown variables is balanced with the number of equations, 
it is not so surprising if one can choose the unknowns so that the objective function $F$ becomes (almost) zero.
However, even if one can obtain the approximated eigenvalue distribution $\rho^{(P)}(\lambda)$ in this way, whether or not it agrees with the loop equations \textit{for all $n$} is a nontrivial issue, and at this stage the validity of the approximate solution is not guaranteed. 
On the other hand, if one starts from an overdetermined system and applies the least-squares method, the very existence of an approximate solution becomes nontrivial. 
In that case, obtaining such a solution strongly suggests that it provides a good approximation to a genuine solution of the loop equations for all $n$. 

Nevertheless, in contrast to the matrix bootstrap approach, this bootstrap approximation does not yield rigorous bounds on physical observables. Therefore, in this approximation scheme, double and triple checks of the approximate solution are always indispensable. Thus, we consider it inadvisable to lower the accuracy of the approximation by adopting an overdetermined system from the outset, and we choose the present setup. Instead, the validity of the approximation is assessed using the test function \(G\)~\eqref{eq:test function G}
together with the positivity constraint, as described in Section~\ref{subsec:Consistency Check of the Approximate Results}.

\subsection{Euclidean-type one-matrix model}

\subsubsection{Endpoint-free approximation}

\begin{figure}[t] 
\centering
\includegraphics[width=0.8\textwidth]{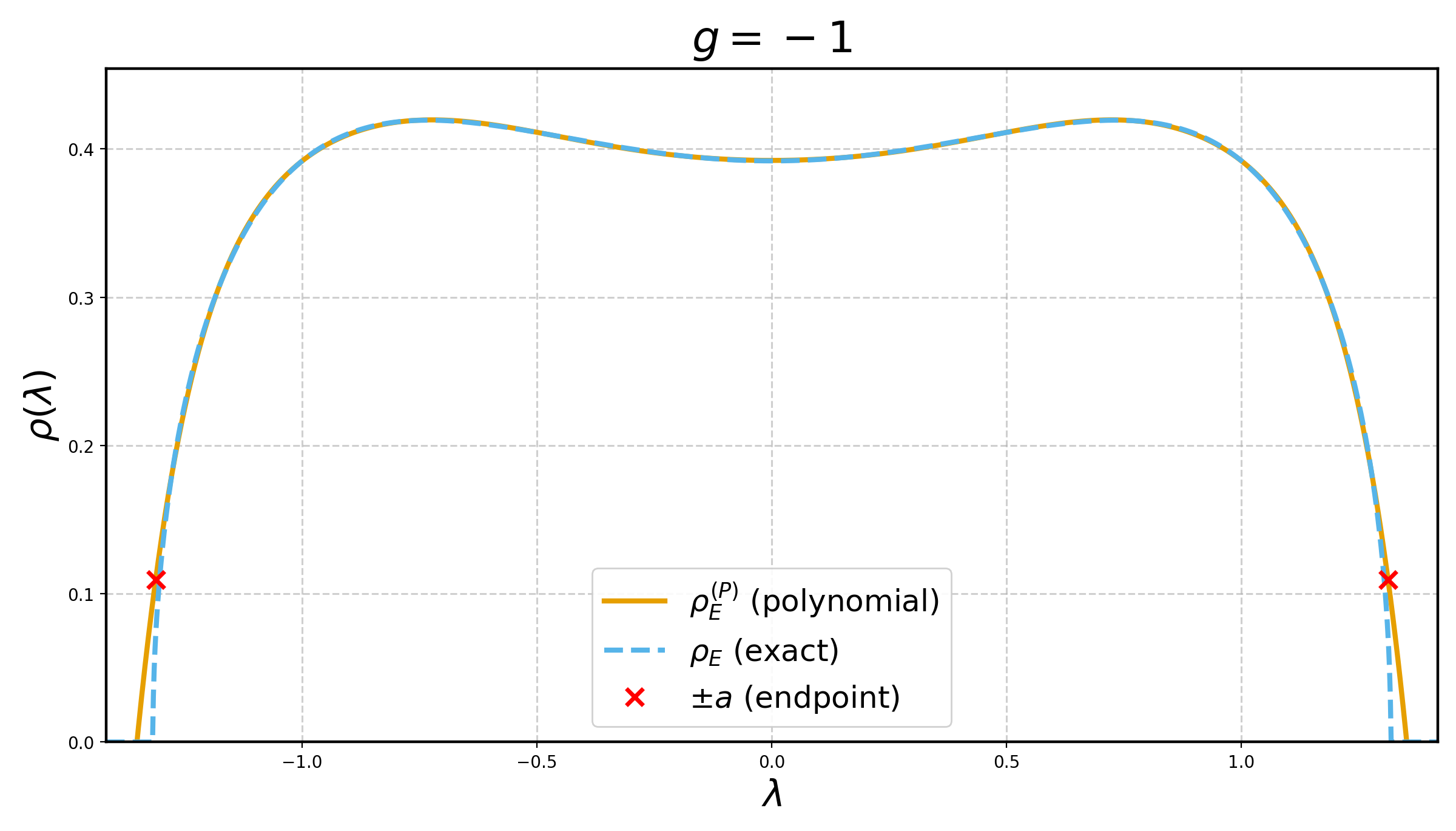}
\hfill
\includegraphics[width=0.8\textwidth]{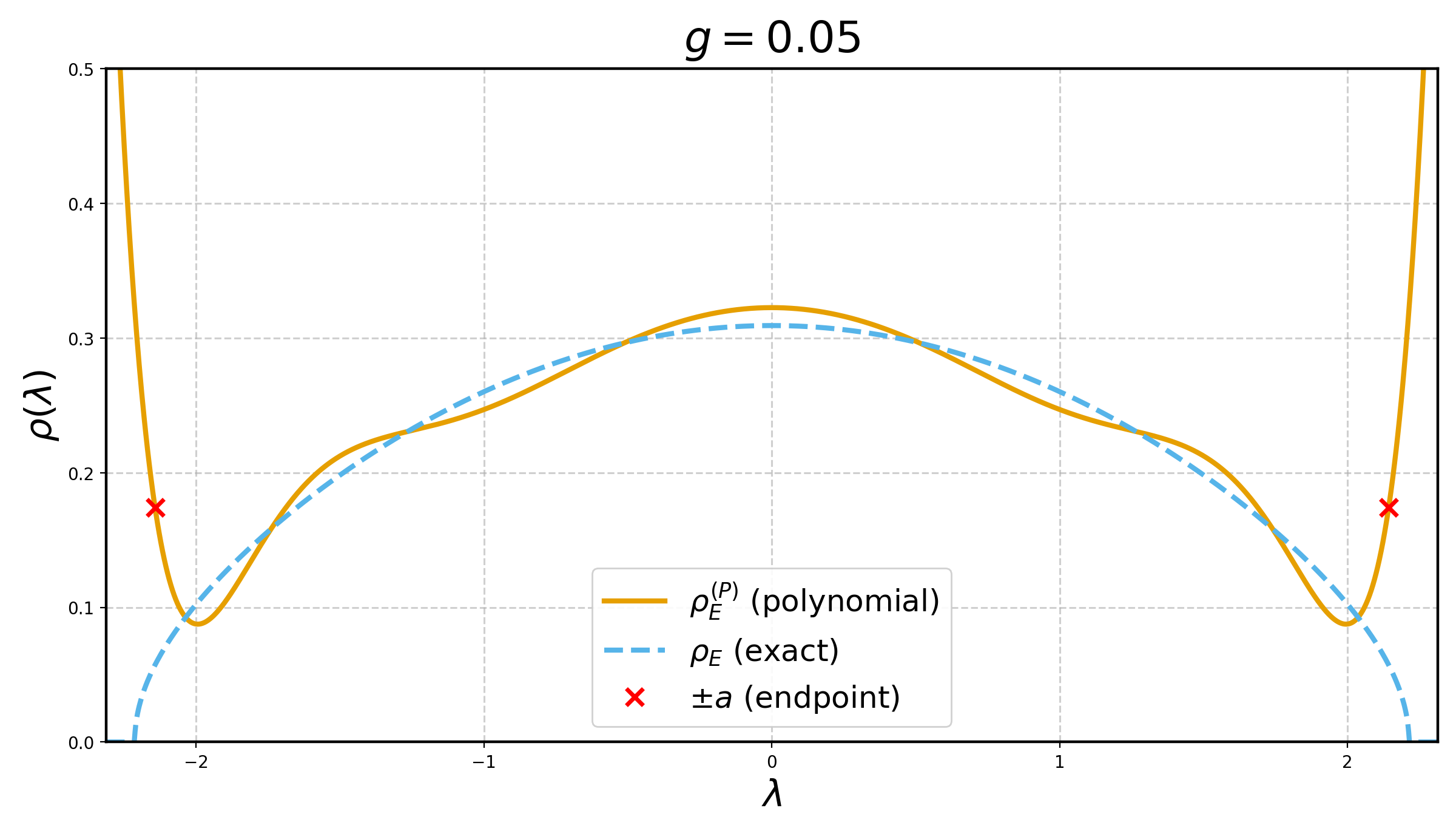}
\caption{
In the two plots, the exact eigenvalue distributions $\rho_{E}(\lambda)$ are compared with its polynomial approximations $\rho_{E}^{(P)}(\lambda)$. 
The upper panel corresponds to $g=-1$, while the lower panel corresponds to $g=0.05$. 
The dashed lines represent $\rho_{E}(\lambda)$, and the solid lines represent $\rho_{E}^{(P)}(\lambda)$. 
In the polynomial approximation, the integration interval is fixed to $[-a,a]$, and the value of $\pm a$ obtained from the bootstrap approximation are indicated by cross marks. 
For both $g=-1$ and $g=0.05$, one finds that $\rho_{E}^{(P)}(\pm a)\neq 0$, and in particular for $g=0.05$ there is a significant discrepancy from the exact solution $\rho_{E}(\pm a)$.
}
\label{fig:dist for g=-1,0.05 without endpoint fixing}
\label{fig:compare}
\end{figure}

\begin{table}[t]
\centering
\begin{tabular}{c c c c c c c}
\hline
$g$ 
& $a^{(P)}$ 
& $w_{2}^{(P)}$
& $\lvert a^{(P)} - a^{\text{exact}}\rvert$
& $\lvert w_{2}^{(P)} - w_{2}^{\text{exact}}\rvert$
& $F_E$ 
& $L$ \\
\hline
$-1$ 
& $1.312$
& $0.5161$
& $5.994\times 10^{-3}$
& $1.572\times 10^{-8}$
& $7.395\times 10^{-31}$
& $173$ \\
\hline
$0.05$ 
& $2.142$
& $1.133$
& $7.075\times 10^{-2}$
& $8.885\times 10^{-8}$
& $1.587\times 10^{-5}$
& $7$ \\
\hline
$0.1$ 
& $1.663$
& $1.772$
& ---
& ---
& $2.940\times 10^{5}$
& $1$ \\
\hline
\end{tabular}
\caption{
This table summarizes the results of the bootstrap approximation for three values of the coupling constant, $g=-1$, $0.05$, and $0.1$.
The quantities $a^{(P)}$ and $w_{2}^{(P)}$ denote the values obtained from the bootstrap approximation, whereas $a^{\text{exact}}$ and $w_{2}^{\text{exact}}$ represent the exact solutions derived using the method described in Appendix~\ref{appendix a}.
For the test function $G_{E}$, $L$ is defined as $L\equiv\Lambda'-\Lambda$, where $\Lambda'$ is the largest value satisfying $G_{E}\le 1$.
Since $g=0.1$ exceeds the critical value $g_{c}=1/12$, the theory is not well defined, and the corresponding values of $|a^{(P)}-a^{\text{exact}}|$ and $|w_{2}^{(P)}-w_{2}^{\text{exact}}|$ are therefore left blank.
}
\label{tab:euclide without EF}
\end{table}

For the bootstrap approximation of the Euclidean one-matrix model, we set the polynomial degree to $M=10$ and choose $\Lambda=14$.
This choice is equivalent to imposing the relations $w_{2n}=w_{2n}^{(P)}$ for $n=0,1,\ldots,7$, yielding a total of eight equations.
For $M=10$, this exactly matches the number of degrees of freedom, namely the eight unknown parameters $c_{0},c_{2},\ldots,c_{10}$ together with $a$ and $w_{2}$.
For the test function $G_{E}$, we make use of the $\mathbb{Z}_{2}$ symmetry and define
\begin{equation}
G_{E}=\sum_{n=8}^{\Lambda'}\frac{|w_{2n}-w_{2n}^{(P)}|^{2}}{|w_{2n}|^{2}}.
\end{equation}
Since the $\mathbb{Z}_{2}$ symmetry implies $w_{2n-1}=0$, it is necessary in numerical computations to skip the odd-order moments in advance in this manner.
As explained in Section~\ref{subsec:Consistency Check of the Approximate Results}, for this test function $G_{E}$ we define $L\equiv\Lambda'-\Lambda$ using the largest $\Lambda'$ that satisfies $G_{E}\le 1$.
A larger value of $L$ can be interpreted as an indication that the approximation is performing well.

The numerical results are summarized in Table~\ref{tab:euclide without EF}.
We examined three values of the coupling constant, $g=-1$, $0.05$, and $0.1$.
Although $g=0.05$ is positive, it still lies in the regime where the theory is well defined, whereas $g=0.1$ belongs to a region in which the theory is no longer well defined (see Appendix~\ref{appendix a} for the critical value $g_{c}=1/12$).
For the objective function $F_{E}$, we find that it approaches zero up to machine precision for $g=-1$, while it takes a somewhat larger value for $g=0.05$, and becomes extremely large for $g=0.1$.
This behavior is considered to strongly reflect the fact that, in the region $g>g_{c}$, the two self-consistency conditions of the present bootstrap approximation—namely, that an eigenvalue distribution exists such that $w_{n}=\int d\lambda\lambda^{n}\rho(\lambda)$ holds, and that the moments $w_{n}$ satisfy the loop equations—can no longer be simultaneously satisfied.

Consistent with this observation, the values of $L$ are significantly large for $g=-1$, while for $g=0.05$ one finds $L=7$, and for $g=0.1$ only $L=1$.
Because odd-order moments can be ignored due to the $\mathbb{Z}_{2}$ symmetry, the result $L=1$ for $g=0.1$ implies that the unknown variables $c_{m},a,w_{2}$ optimized using the information up to $w_{0},\dots,w_{14}$ are completely ineffective in reducing $|w_{16}-w_{16}^{(P)}|$.
In other words, one may conclude that these variables $c_{m},a,w_{2}$ do not provide an algebraic solution to the loop equations.
On the other hand, judging from the values of $F_{E}$ and $L$, the result for $g=0.05$ should be regarded as somewhat marginal.
While $L=7$ suggests the existence of a nontrivial algebraic solution to the loop equations, this value is rather modest compared with $L=173$ obtained for $g=-1$, and the value $F_{E}\sim10^{-5}$ also indicates that the optimization of the bootstrap approximation for $g=0.05$ has not been fully successful.
This interpretation will be fully justified in Figure~\ref{fig:dist for g=-1,0.05 without endpoint fixing}.

However, even for $g=0.05$, the quantity $|w_{2}^{(P)}-w_{2}^{\text{exact}}|$ becomes as small as in the case $g=-1$.
This implies that, for low-order moments, the approximation is extremely accurate, despite the fact that the optimization is not fully successful.
Our current view on this seemingly puzzling phenomenon is as follows.

As already discussed in Section~\ref{subsec:Consistency Check of the Approximate Results}, higher-order moments $w_{n}$ behave asymptotically as $a^{n}$, whereas lower-order moments, such as $w_{2}$ and $w_{2}^{(P)}$ themselves, contribute only very weakly to the objective function $F_{E}$.
Nevertheless, through the loop equations, $w_{2}$ determines all higher moments $w_{4},w_{6},\dots$.
In particular, since higher-order moments $w_{n}$ take extremely large values, even a slight deviation of $w_{2}$ from its ``ideal'' value leads to an explosive growth of $|w_{n}-w_{n}^{(P)}|^{2}$.
Therefore, in order to reduce the value of the objective function $F_{E}$, the least-squares procedure must act so as to finely tune the value of $w_{2}$, forcing it to agree with the ``ideal'' value with high precision.
We consider this ``ideal'' value to be $w_{2}^{\text{exact}}$.
Indeed, the current bootstrap approximation is formulated based on the self-consistency condition that the existence of an eigenvalue distribution $\rho_{E}(\lambda)$ is compatible with the loop equations, and under this condition the allowed value of $w_{2}$ should be uniquely fixed to $w_{2}^{\text{exact}}$.

In summary, requiring $|w_{n}-w_{n}^{(P)}|^{2}$ to be small for all $n$ automatically induces a strong fine-tuning that forces $w_{2}$ to coincide with $w_{2}^{\text{exact}}$, which presumably originates from the self-consistency conditions of the bootstrap approximation.
From a practical point of view, the fact that lower-order moments tend to be better approximated can be regarded as an advantage.
At the same time, as explained above, $|w_{n}-w_{n}^{(P)}|^{2}$ naturally grows for large $n$.
Therefore, when higher-order moments are included in the approximation, namely when a larger cutoff $\Lambda$ is used, choosing appropriately smaller weights $r_{n}$ in the objective function $F_{E}$ may lead to a better approximation.

In addition to Table~\ref{tab:euclide without EF}, in the Euclidean model, the exact eigenvalue distribution $\rho_{E}(\lambda)$ is known in the region $g<g_{c}$.
Therefore comparing it with its polynomial approximation $\rho_{E}^{(P)}(\lambda)$ is useful for assessing the performance of the present method. 
Accordingly, in Figure~\ref{fig:dist for g=-1,0.05 without endpoint fixing} we present a comparison between $\rho_{E}(\lambda)$ and $\rho_{E}^{(P)}(\lambda)$ for $g=-1$ and $g=0.05$. 
The upper panel corresponds to $g=-1$, while the lower panel corresponds to $g=0.05$. 

For $g=-1$, the quality of the fit of $\rho_{E}^{(P)}(\lambda)$ is remarkable even up to the vicinity of the endpoints $\pm a$, to the extent that it is almost impossible to distinguish the two curves. 
One should note, however, that the present $\rho_{E}^{(P)}(\lambda)$ does not vanish at the endpoints $\pm a$, and instead satisfies $\rho_{E}^{(P)}(\pm a) > 0$. 
This behavior is presumably due to the fact that no mechanism enforcing $\rho_{E}^{(P)}(\pm a)=0$ was incorporated into the polynomial approximation~\eqref{eq: approximation anzats}. 

In contrast, the plot for $g=0.05$ clearly indicates that the approximation does not work so well. 
Although one may still argue that the fit is acceptable up to around $\lambda=\pm 2$, outside this region $\rho_{E}^{(P)}(\lambda)$ increases rather than decreases, exhibiting behavior that is far removed from the true eigenvalue distribution.

\subsubsection{Endpoint-fixed approximation}

\begin{table}[t]
\centering
\begin{tabular}{c c c c c c c}
\hline
$g$ 
& $a^{(P)}$ 
& $w_{2}^{(P)}$
& $\lvert a^{(P)} - a^{\text{exact}}\rvert$
& $\lvert w_{2}^{(P)} - w_{2}^{\text{exact}}\rvert$
& $F_E$ 
& $L$ \\
\hline
$-1$ 
& $1.330$
& $0.5161$
& $1.224\times 10^{-2}$
& $5.932\times 10^{-7}$
& $5.007\times 10^{-30}$
& $89$ \\
\hline
$0.05$ 
& $2.234$
& $1.133$
& $2.104\times 10^{-2}$
& $2.351\times 10^{-8}$
& $5.137\times 10^{-22}$
& $15$ \\
\hline
\end{tabular}
\caption{
This table summarizes the results of the bootstrap approximation for $g=-1$ and $0.05$ with the endpoint fixed.
}
\label{tab:euclide with EF}
\end{table}

\begin{figure}[t] 
\centering
\includegraphics[width=0.8\textwidth]{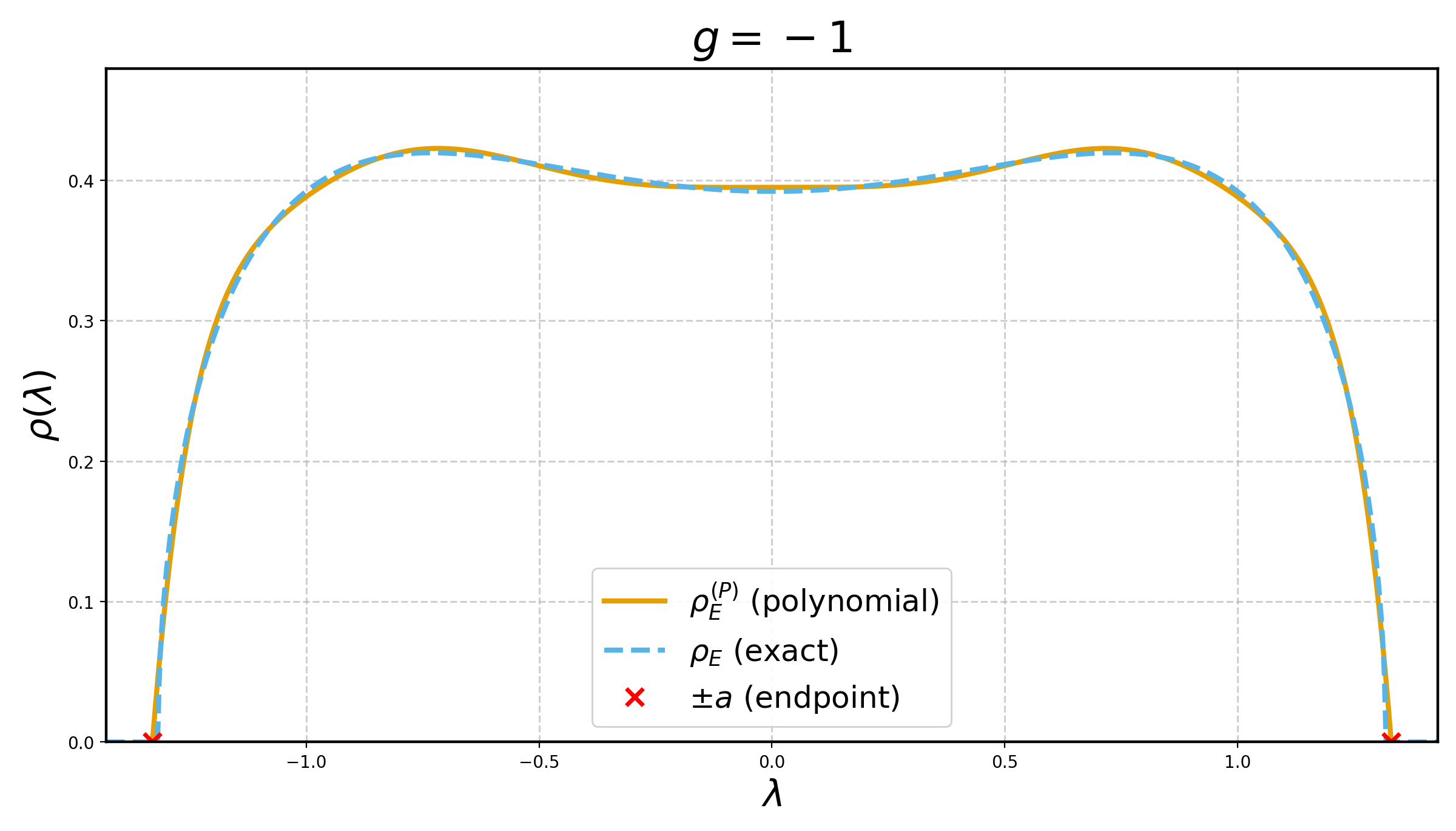}
\hfill
\includegraphics[width=0.8\textwidth]{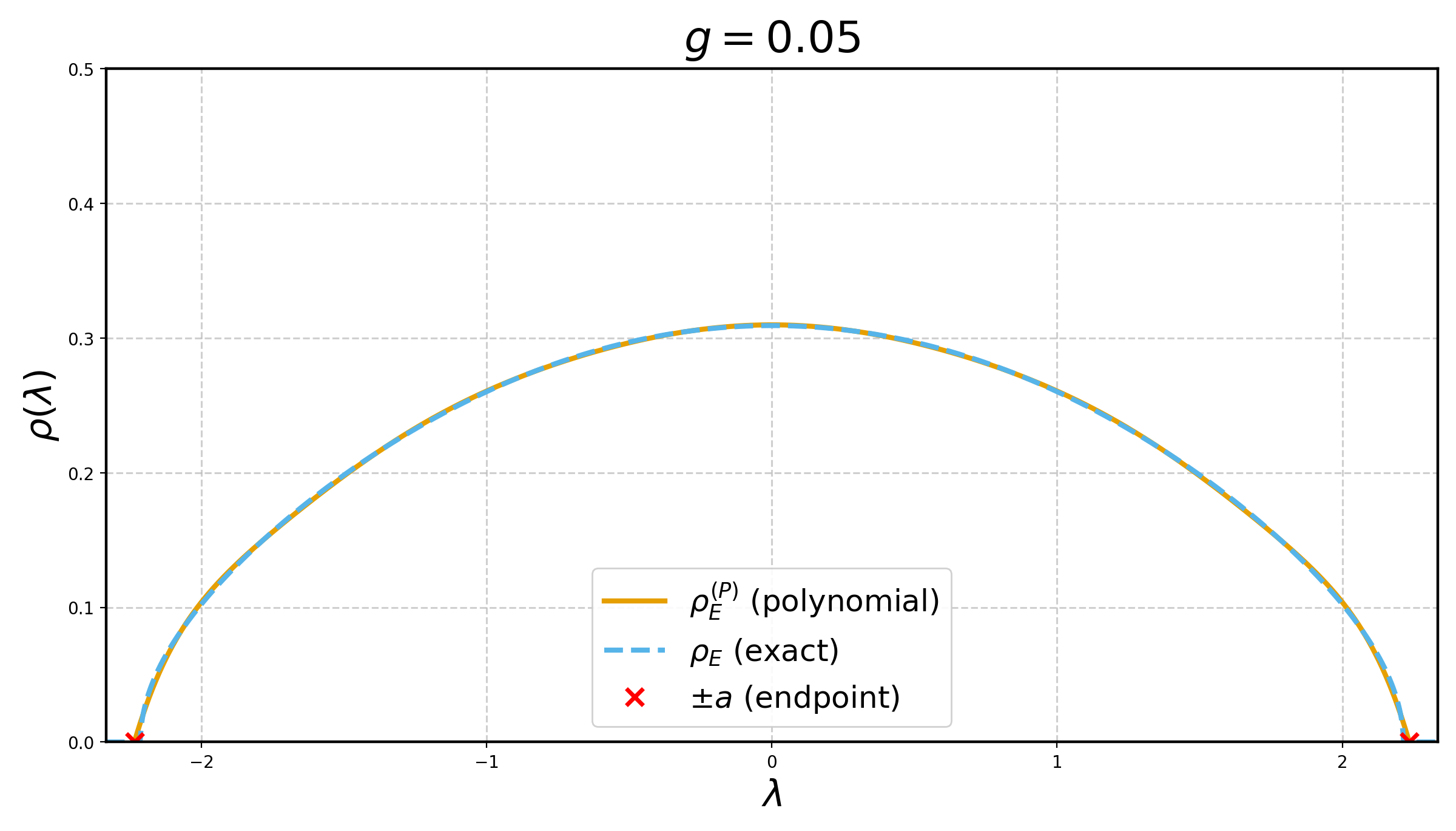}
\caption{
These figures show $\rho_{E}^{(P)}(\lambda)$ obtained by imposing the condition $\rho_{E}^{(P)}(\pm a)=0$ in~\eqref{eq:new approximation anzats}. 
In this case, a high level of accuracy is achieved for both $g=-1$ and $g=0.05$. 
As expected, the condition $\rho_{E}^{(P)}(\pm a)=0$ is indeed satisfied at the endpoints $\pm a$, which are indicated by cross marks.
}
\label{fig:dist for g=-1,0.05 with endpoint fixing}
\end{figure}

As already seen most clearly in the case $g=0.05$, the approximation scheme introduced above tends to exhibit unstable behavior near the endpoints $\pm a$, which may be responsible for the degradation of the approximation accuracy. To remedy this issue, we now consider a modified ansatz in which the polynomial approximation $\rho_{E}^{(P)}(\lambda)$ is taken to be
\begin{align} \label{eq:new approximation anzats}
\rho_{E}^{(P)}(\lambda)
= (\lambda-a)(\lambda+a)\sum_{m=0}^{M/2-1} c_{2m}\lambda^{2m}.
\end{align}
With this choice, the condition $\rho_{E}^{(P)}(\pm a)=0$ is enforced by construction. It should be noted that, compared with the endpoint-unfixed case, this ansatz reduces the number of independent coefficients $c_{2m}$ by one. Consequently, even when considering a polynomial of total degree ten, the number of unknown variables is reduced from eight to seven. In accordance with this reduction, we also decrease the number of imposed equations by setting $\Lambda=12$.

The resulting eigenvalue distributions for $g=-1$ and $g=0.05$ are shown in Figure~\ref{fig:dist for g=-1,0.05 with endpoint fixing}. In particular, the dramatic improvement in the approximation accuracy for $g=0.05$ is noteworthy. This clearly demonstrates that the loss of accuracy observed in the endpoint-unfixed approximation for $g=0.05$ originates from the behavior near the endpoints $\pm a$.
Table~\ref{tab:euclide with EF} also summarizes the values of the unknown variables obtained in the approximation, together with the value of the objective function.

Note that, in the endpoint-fixed approximation, the accuracy for $g=-1$ slightly deteriorates, and accordingly $|w_{2}^{(P)}-w_{2}^{\text{exact}}|$ becomes somewhat larger.
Presumably reflecting this effect, the value of $L$ also decreases slightly to $L=89$.
Nevertheless, as can be seen from Figure~\ref{fig:dist for g=-1,0.05 with endpoint fixing}, the fit of the eigenvalue distribution remains good.
Furthermore, for $g=0.05$, the value of $L$ improves compared with the endpoint-free approximation, yielding $L=15$.
Although this may appear small when compared with the case $g=-1$, taking into account the fact that the eigenvalue distribution is well fitted, the value $L=15$ should not be regarded as poor.
Rather, it is more appropriate to interpret the value $L\sim100$ as being exceptionally large.

\begin{table}[t]
\centering
\setlength{\tabcolsep}{4pt}
\begin{tabular}{c c c c}
\hline
$g$ & approximation & $w_2$ &
sign of $\det \bar H^{(k)}$ for $k=1,\dots,30$ \\ \hline
$-1$ &
exact &
0.5161512329820714 &
\texttt{+++++|+++++|+++++|+++++|+++--|-----} \\

$-1$ &
endpoint-free &
0.5161512172520892 &
\texttt{+++++|+++++|+----|-----|-----|+++++} \\

$-1$ &
endpoint-fixed &
0.5161506397441684 &
\texttt{+++++|++++-|-----|-----|-++++|+++++} \\ \hline
$0.05$ &
exact &
1.133201576396077 &
\texttt{+++++|+++++|+++++|+++++|+++++|+++++} \\

$0.05$ &
endpoint-free &
1.133201487536680 &
\texttt{+++++|+++-+|+++++|+++-+|+++++|++++-} \\

$0.05$ &
endpoint-fixed &
1.133201599909633 &
\texttt{+++++|+++++|+++++|++++-|+++++|+++++} \\ \hline
$0.1$ &
endpoint-free &
1.772103894787935 &
\texttt{+++++|-++++|++-++|++++-|+++++|+-+++} \\ \hline
\end{tabular}
\caption{
Sign patterns of $\det[\bar{H}^{(k)}(w_{2}^{(P)})]$ for $k=1,\dots,30$.
The delimiter \texttt{|} is inserted every five entries for readability.
From top to bottom, the values of $K$ are 23, 11, 9, at least 30, 8, 19, and 5.
}
\label{tab:positivity_sign_patterns}
\end{table}

Finally, for both the endpoint-free and endpoint-fixed cases, we performed the consistency check based on positivity introduced in Section~\ref{subsec:Consistency Check of the Approximate Results}.
The results are summarized in Table~\ref{tab:positivity_sign_patterns}.
Here, numerical inputs are taken up to the 16th decimal place, and the numerical precision is set to 1000 digits.
For completeness, we have examined the sign of $\det[\bar{H}^{(k)}(w_{2}^{(P)})]$ for all $k=1,...,30$ in every case,
but the quantity of primary interest is $K$ satisfying
\begin{equation}
\forall k\in\{1,...,K\},\quad 
\ \det[\bar{H}^{(k)}(w_{2}^{(P)})]>0,\ \quad \det[\bar{H}^{(K+1)}(w_{2}^{(P)})]<0
\end{equation}
(which is equal to the maximal $k$ for which $\bar{H}^{(k)}(w_{2}^{(P)})\succcurlyeq0$ holds).
As expected, for both $g=-1$ and $g=0.05$, $w_{2}^{\text{exact}}$ yields the largest value of $K$.
Moreover, in both the endpoint-free and endpoint-fixed cases,
values of $w_{2}^{(P)}$ closer to $w_{2}^{\text{exact}}$ correspond to larger $K$,
indicating that even such small differences in numerical values are correctly reflected in the results.
By contrast, for $g=0.1$, where positivity is violated,
the value of $w_{2}^{(P)}$ obtained from the endpoint-free approximation gives only $K=5$,
from which the failure of the approximation can be clearly inferred.

In this way, the verification of positivity constraints serves as a nontrivial consistency check.
Nevertheless, it should also be noted that this method is not perfect.
In particular, since the bootstrap approximation employed in this work tends to be more accurate for lower-order moments, the positivity-based inequalities are likely to be satisfied with high probability as long as one tests only $w_{2}^{(P)}$.
However, as seen in the free-endpoint case at $g=0.05$, even if $w_{2}^{(P)}$ is well approximated, the polynomial approximation of the eigenvalue distribution $\rho_{E}^{(P)}(\lambda)$ does not necessarily reproduce the exact distribution $\rho_{E}(\lambda)$ accurately.
Such failures of the approximation are often invisible when one inspects only $w_{2}^{(P)}$, and become apparent only after examining higher-order moments.
Fortunately, the quantities $F_{E}$ and $L$ introduced above incorporate information from these higher-order moments.
At present, therefore, it appears to be a reasonable compromise to assess the quality of the approximation by appropriately combining these indicators.

\subsection{Minkowski-type one-matrix model}

As discussed in Section~\ref{sebsec:overall setup}, in the Minkowski-type one-matrix model an additional angular (phase) parameter $\theta$ appears, and the coefficients $c_{2m}$ as well as the moments $w_{2n}$ become complex-valued. 
In the following, we adopt the endpoint-fixed ansatz~\eqref{eq:new approximation anzats} from the outset and, as in the Euclidean case, assume a polynomial of degree ten. In this setup, the unknown variables are $c_{0},\ldots,c_{8},\theta,a,$ and $w_{2}$, which amount to $14$ real degrees of freedom. We therefore set $\Lambda=12$ and incorporate the conditions
$w_{0}=w_{0}^{(P)}, w_{2}=w_{2}^{(P)}, \ldots, w_{12}=w_{12}^{(P)}$
into the least-squares procedure. In this way, the number of equations matches the number of unknowns. Note that both $w_{n}$ and $w_{n}^{(P)}$ are complex, and each condition $w_{n}=w_{n}^{(P)}$ thus provides two real equations. 
Unlike the Euclidean model, we set $r_{n}=(0.8)^n$ in the present analysis in order to improve numerical stability.

Here, as a ``formal eigenvalue distribution'' of the Minkowski-type one-matrix model, we define $\rho_{M}(z)$ following the discussion in Appendix~\ref{appendix b}. 
Our stance on this point is as follows. 
First, when the value of the test function $G$ is sufficiently small—especially when it is comparable to or even smaller than in the Euclidean case—the present approximation is likely to provide an algebraic solution of the loop equations. 

Furthermore, since the formal eigenvalue distribution $\rho_{M}(z)$ reproduces the perturbative expansion, if our approximation succeeds in reproducing this $\rho_{M}(z)$ with sufficiently small value of $g$, it strongly suggests that the approximation also captures a ``physical'' solution of the loop equations in such a small $g$ region.
However, we emphasize once again that the perturbative expansion itself cannot be trusted for large values of $g$.
Therefore, even if the optimization appears to work well in that regime, the resulting approximate values cannot immediately be regarded as physically meaningful.
In the present paper we will not pursue the physical interpretation of the large-$g$ regime any further, and leave a detailed analysis to a forthcoming paper.

\begin{table}[t]
\centering
\small
\setlength{\tabcolsep}{3pt}
\begin{tabular}{c c c c c c c}
\hline
$g$
& $w_{2}^{(P)}$
& $\lvert a^{(P)} - a^{\text{formal}}\rvert$
& $\lvert \theta^{(P)} - \theta^{\text{formal}}\rvert$
& $\lvert w_{2}^{(P)} - w_{2}^{\text{formal}}\rvert$
& $F_M$
& $L$ \\
\hline
$-1$
& $0.2776 + 0.4934\,i$
& $1.272\times 10^{-2}$
& $7.203\times 10^{-5}$
& $1.367\times 10^{-6}$
& $2.557\times 10^{-30}$
& $89$ \\
\hline
$1$
& $-0.2776 + 0.4934\,i$
& $1.272\times 10^{-2}$
& $7.203\times 10^{-5}$
& $1.367\times 10^{-6}$
& $4.653\times 10^{-29}$
& $89$ \\
\hline
$-0.3$
& $0.2695 + 0.7555\,i$
& $1.519\times 10^{-1}$
& $7.985\times 10^{-4}$
& $2.396\times 10^{-6}$
& $1.272\times 10^{-28}$
& $25$ \\
\hline
$-0.1$
& $0.1635 + 0.9339\,i$
& $1.799\times 10^{-2}$
& $6.300\times 10^{-5}$
& $2.424\times 10^{-8}$
& $2.253\times 10^{-24}$
& $17$ \\
\hline
\end{tabular}
\caption{
Bootstrap approximations for the Minkowski-type one-matrix model performed with $M=10$, $\Lambda=12$, $r_n=(0.8)^n$, and $g=\pm1,-0.3,-0.1$.
The quantities $a^{\text{formal}}$ and $\theta^{\text{formal}}$ are computed according to the formal solution given in Appendix~\ref{appendix b}.
}
\label{tab:minkowski_EF}
\end{table}

\begin{figure}[t] 
\centering
\includegraphics[width=0.9\textwidth]{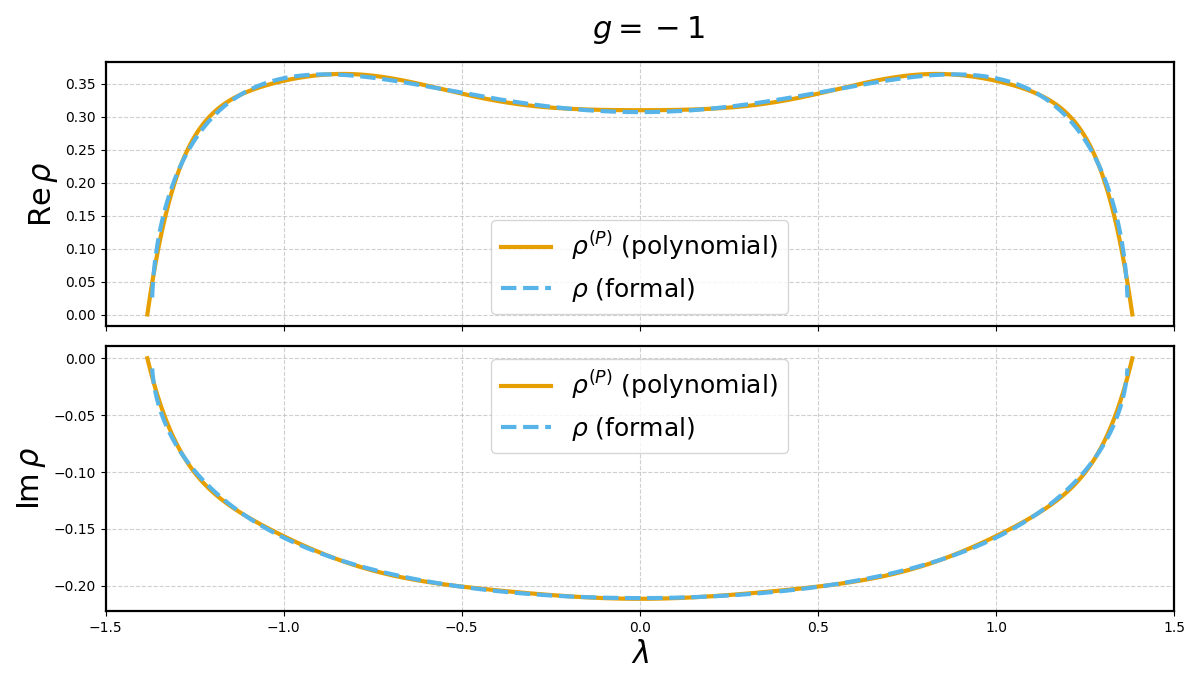}
\hfill
\includegraphics[width=0.9\textwidth]{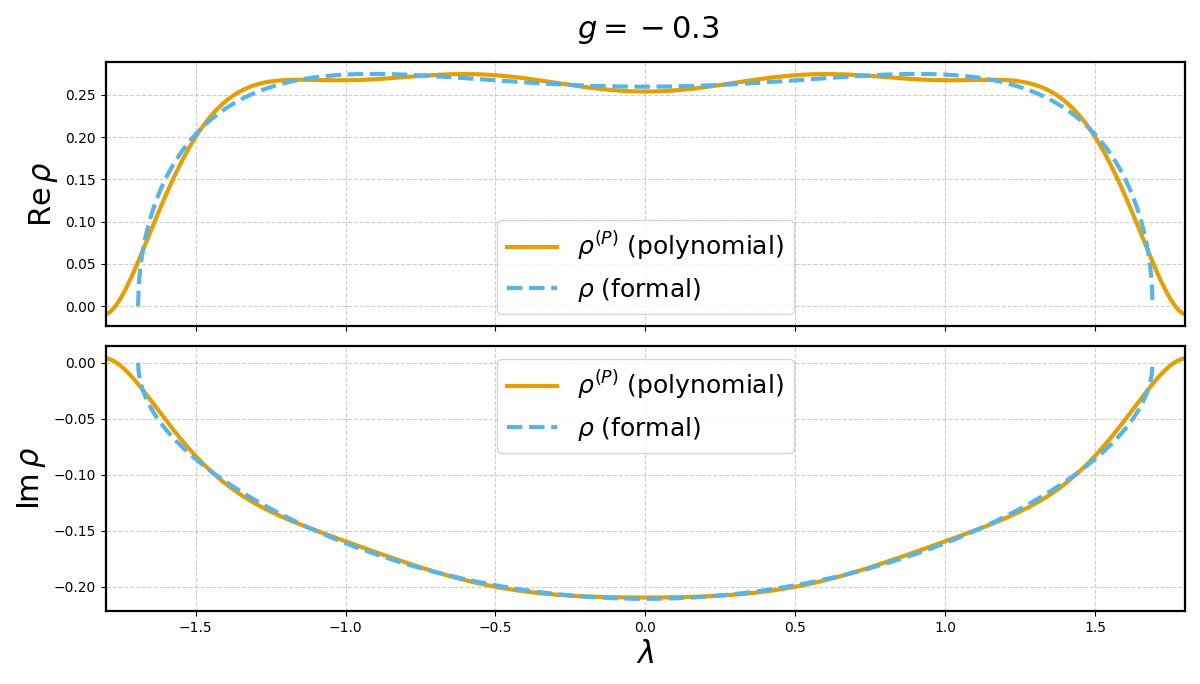}
\caption{
Plots of the real and imaginary parts of the formal solution $\rho_M(z)$ and its polynomial approximation $\rho_M^{(P)}(z)$ for $g=-1$ and $g=-0.3$.
Each of $\rho_M(z)$ and $\rho_M^{(P)}(z)$ is defined as a complex-valued function supported on $\Gamma$ and $\Gamma^{(P)}$, respectively.
By parametrizing them as $\rho_M(e^{i\theta}\lambda)$ and $\rho_M^{(P)}(e^{i\theta^{(P)}}\lambda)$, the two distributions are plotted on the same $\lambda$--$\rho$ plane.
Note, however, that as a consequence of this parametrization, the difference between the angles $\theta$ and $\theta^{(P)}$ cannot be read off from this figure (however, as can be seen from Table~\ref{tab:minkowski_EF}, the numerical difference between these quantities is in fact extremely small).
}
\label{fig:M_dist for g=-1}
\end{figure}

With these considerations in mind, Figure~\ref{fig:M_dist for g=-1} presents  comparisons between $\rho_M^{(P)}(z)$ obtained from the bootstrap approximation and the formal solution $\rho_M(z)$ with $g=-1$ (top panel) and $g=-0.3$ (bottom panel).
Both distributions are defined on line segments in the complex plane connecting $\pm e^{i\theta} a$.
Here it should be emphasized that the parameters $\theta$ and $a$ are in general different for $\rho_M(z)$ and $\rho_M^{(P)}(z)$.
To avoid confusion, we denote by $\theta$ and $a$ the parameters associated with $\rho_M(z)$, whose values are determined by~\eqref{eq:formal solution of alpha}, while $\theta^{(P)}$ and $a^{(P)}$ are those determined by the bootstrap approximation.
With this notation, the supports can be written explicitly as
\begin{equation}
\begin{aligned}
\Gamma & = \{ e^{i\theta} \lambda \mid \lambda \in [-a,a] \}, \\
\Gamma^{(P)} & = \{ e^{i\theta^{(P)}} \lambda \mid \lambda \in [-a^{(P)},a^{(P)}] \}.
\end{aligned}
\end{equation}
Accordingly, $\rho_M(z)$ and $\rho_M^{(P)}(z)$ are defined on $\Gamma$ and $\Gamma^{(P)}$ as $\rho_M(e^{i\theta}\lambda)$ and $\rho_M^{(P)}(e^{i\theta^{(P)}}\lambda)$, respectively.
When plotting the comparison between the two distributions, we evaluate $\rho_M(z)$ and $\rho_M^{(P)}(z)$ on their respective supports $\Gamma$ and $\Gamma^{(P)}$, and then display them on the same $\lambda$--$\rho$ plane.
As a consequence, the difference between the angles $\theta$ and $\theta^{(P)}$ cannot be read off directly from Figure~\ref{fig:M_dist for g=-1} (however, as can be seen from Table~\ref{tab:minkowski_EF}, the numerical difference between these quantities is in fact extremely small).
The numerical values of the relevant parameters are summarized in Table~\ref{tab:minkowski_EF}.
Since the validity of the results cannot be guaranteed for large absolute values of $g$, we present here only four cases, $g=\pm1,-0.3,-0.1$.
However, we note that, based on our actual numerical computations, the optimization itself converges successfully even in the region $|g|>1$, with performance comparable to that for $|g|<1$.

From these figures, one can see that, especially for $g=-1$, our approximation method reproduces the formal eigenvalue distribution $\rho_{M}(z)$ extremely well.
The value of $L$ is at least $L=17$, which can be regarded as sufficiently large when compared with the value $L=15$ obtained for the Euclidean model at $g=0.05$ in the endpoint-fixed approximation.
Also, as already discussed in the Euclidean case, this method tends to yield higher accuracy for lower-order moments, which is confirmed by the fact that $|w_{2}^{(P)}-w_{2}^{\text{formal}}|$ is of order less than $10^{-6}$ for all values of $g$.
Overall, the method can be evaluated as functioning as expected also for the Minkowski-type model.

However, for $g=-0.3,-0.1$, the objective function $F_{M}$ does not become sufficiently small.
Technically, this is because, at a certain stage of the optimization process, xtol rapidly becomes very small and the approximation stops at that point.
As a result, as shown in Figure~\ref{fig:M_dist for g=-1}, the eigenvalue distribution is not completely fitted.
From a theoretical viewpoint, however, this situation is likely different from the Euclidean case at $g=0.1$, where the theory does not possess a physical solution.
Indeed, the formal eigenvalue distribution $\rho_{M}(z)$ exactly reproduces the perturbative expansion in $g$, and there is no apparent reason for the physical solution to break down in the small-$|g|$ region.\footnote{In the Euclidean case, even for $|g|\ll1$, it is intuitively expected that the matrix integral becomes ill-defined for $g>0$.
In this sense, the existence of a physical solution for $0<g<g_{c}$ is rather nontrivial.}
Our conjecture regarding this issue is that, as explained in Appendix~\ref{appendix b}, the formal solution of the Minkowski-type one-matrix model requires different branch choices depending on the sign of $g$, and that this leads to numerical instabilities.

We add several further comments on the numerical results.  
First, as shown in Figure~\ref{fig:M_dist for g=-1}, in the present setup the imaginary parts of the formal eigenvalue distribution $\rho_{M}(z)$ and its bootstrap approximation $\rho_{M}^{(P)}(z)$ are negative on their support, which may appear to contradict the fact that $\operatorname{Im} w_{2}>0$.  
However, this intuition is in fact incorrect.  
This is because we defined $w_{n}$ and $\rho_{M}(z)$ so that $w_{n}=\int_{\Gamma}dz\,z^{n}\rho_{M}(z)$ holds, and therefore, when the support takes the form $\Gamma=\{e^{i\theta}\lambda \mid \lambda\in[-a,a]\}$, $w_{2}$ is computed as
\begin{equation}
w_{2}=\int_{-a}^{a}d(e^{i\theta}\lambda)(e^{i\theta}\lambda)^{2}\rho_{M}(e^{i\theta}\lambda)
= e^{3i\theta}\int_{-a}^{a}d\lambda\,\lambda^{2}\rho_{M}(e^{i\theta}\lambda).
\end{equation}
The eigenvalue-distribution plots are obtained simply by evaluating $\rho_{M}(e^{i\theta}\lambda)$ on its support $\Gamma$, and there is a phase difference given by the factor $e^{3i\theta}$.

The next comment concerns the results for $g=1$.  
As is immediately seen from Table~\ref{tab:minkowski_EF}, the value of $w_{2}$ for $g=1$ differs from that for $g=-1$ only by a sign flip of the real part. 
This behavior can be understood intuitively from the perturbative expansion of $w_{2}$: $w_{2}=i-2g-9ig^{2}+\cdots$.  
According to this expansion, the real part of $w_{2}$ is composed only of odd powers of $g$, while the imaginary part is composed only of even powers of $g$.  
Therefore, as long as one trusts this perturbative expansion, changing the sign of $g$ flips only the sign of the real part of $w_{2}$.

Finally, we comment on the phase transition of the Minkowski model.  
In the Euclidean model, there exists a critical value $g_{c}$ of $g$, beyond which a phase transition occurs and the theory breaks down.  
In the bootstrap approximation, this fact is presumably reflected in the failure of the optimization at $g=0.1$, even though the number of unknown variables is balanced with the number of equations.  
In contrast, in the Minkowski model, as far as we have investigated, no such behavior is observed, except for the instability around $g=0$.  
However, as we have repeatedly emphasized, the ansatz employed in this work essentially relies on the one-cut assumption, and in that sense it lacks generality.
Even if one were able to remove this assumption and carry out the optimization, it is not clear whether the same behavior would still be observed.
Therefore it would be premature at this stage to draw any definite conclusion regarding the absence of a phase transition.

In the case of the Euclidean one-matrix model, the existence of a critical coupling $g_{c}$ played an essential role in taking the double-scaling limit and establishing its equivalence with Liouville theory.
In view of this physical importance, the possible existence of a phase transition in the Minkowski one-matrix model clearly deserves further investigation.
For detailed calculations and discussions, see Appendix~\ref{appendix b}.

\section{Summary and Discussion} \label{sec:Summary and Discussion}

In numerical studies of matrix models, direct evaluations of matrix integrals using Monte Carlo methods have been extensively carried out, greatly advancing our understanding of theories primarily at finite $N$.
On the other hand, within Monte Carlo approaches, taking the limit $N \to \infty$ is not possible in principle.
From a theoretical point of view, however, the large-$N$ limit is expected to drastically simplify the structure of the system, while analytical calculations are in fact much more difficult at finite $N$, so this represents a rather curious inversion.
Only recently has this situation begun to change, thanks to the development of matrix bootstrap methods based on semidefinite programming (SDP).
Turning to our approach, we make more active use of the fact that large-$N$ matrix models are expected to possess a simple underlying structure, and translate this structure more directly into a numerical framework.
One of the benefits of this approach is that the simplification due to the large-$N$ limit also extends to Minkowski-type matrix models, which makes it possible to apply the present method to the Minkowski-type one-matrix model under certain assumptions.

Concretely, in this work we have proposed a bootstrap approximation method based on the eigenvalue distribution $\rho(\lambda)$ and applied it to both Euclidean- and Minkowski-type one-matrix models.
The method is designed to determine the (polynomially approximated) eigenvalue distribution $\rho^{(P)}(\lambda)$ and the values of the moments $w_1, w_2$ self-consistently so as to satisfy the following three conditions:
(i) there exists a probability distribution $\rho(\lambda)$ that generates the moments $w_n$;
(ii) $\rho(\lambda)$ is a continuous function with finite support and can be well approximated by a finite-order polynomial;
and (iii) the moments $w_n$ obey the loop equations.
In practice, this is achieved by implementing a least-squares method that minimizes the objective function $ F = \sum_{n}^{\Lambda} \lvert w_n - w_n^{(P)} \rvert^2$ .
In addition, in order to apply this method to Minkowski-type one-matrix models, we have formally extended the notion of the eigenvalue distribution by relying on the large-$N$ factorization and the large-$N$ master-field conjecture.  
Under this assumption, the method can be applied to both Euclidean- and Minkowski-type models on the same footing.  
In fact, the numerical results agree extremely well with the exact solution in the Euclidean case and with the perturbative results in the Minkowski case.

The most important next step of this research is to apply the present method to more complicated matrix models involving multiple types of matrices. 
Of particular interest is the application of the present method to the IKKT matrix model, which is regarded as a nonperturbative formulation of type~IIB superstring theory, as well as to the (Twisted) Eguchi--Kawai model, known as the large-$N$ reduced version of the $U(N)$ gauge theory.
A major challenge in multi-matrix models is the presence of various types of moments, such as $\langle \operatorname{Tr} A^{2} \rangle$, $\langle \operatorname{Tr} B^{2} \rangle$, and $\langle \operatorname{Tr} AB \rangle$, which are generated from different \textit{resolvents}~\eqref{eq:resolvent}. 
For the present method to work, the number of unknown variables to be determined numerically must remain smaller than the number of constraints imposed by the loop equations. However, increasing the number of loop equations inevitably introduces additional types of moments, which in turn require introducing additional resolvents, or eigenvalue distributions, as unknowns. This leads to a deadlock, as the number of unknown variables grows together with the number of constraints.

Our proposal to this problem is that, instead of relying on the eigenvalue distributions, one should make use of the master fields $\hat{A}_{\mu}$ and the density matrix $\hat{\rho}$ introduced in Section~\ref{subsubsec:density-matrix interpretation}.
In fact, as far as multi-matrix models are concerned, the eigenvalue distributions are not particularly powerful, at least for the purpose of computing physical observables.
Even in the case of the simplest nontrivial example beyond the one-matrix model, namely the two-matrix model, infinitely many distinct eigenvalue distributions are required in order to reproduce all possible moments.
However, recalling the arguments based on large-$N$ factorization and the large-$N$ master field in Section~\ref{subsubsec:density-matrix interpretation}, it is clear that these infinitely many eigenvalue distributions are highly redundant.
Indeed, if the large-$N$ master field conjecture~\eqref{eq:master field 2} is true, then a finite number of operators $\hat{A}_{\mu}$ and $\hat{\rho}$ (or equivalently $\bar{A}_{\mu}$) already contain the complete information of the system.

If the master fields and the density matrix actually exist, then by inserting an appropriate resolution of the identity $1=\int dx\,|x\rangle\langle x|$, one can represent $\hat{A}_{\mu}$ and $\hat{\rho}$ as two-variable functions $A_{\mu}(x,y)=\langle x|\hat{A}_{\mu}|y\rangle$ and $\rho(x,y)=\langle x|\hat{\rho}|y\rangle$.
The remaining task is then to approximate these functions by polynomials.
An advantage of this approach is that, unlike the strategy of directly approximating the eigenvalue distributions themselves, the computation safely closes with just a finite number of functions.
As a caveat, however, the action of operators such as $\hat{A}_\mu$ and $\hat{\rho}$ can sometimes be singular and incompatible with a polynomial description, much like in quantum mechanics where $\hat{p}\lvert q\rangle = i\,\partial_q \lvert q\rangle$.
Moreover, as is often the case for the Wigner function in quantum mechanics, it is quite possible that $A_\mu(x,y)$ and $\rho(x,y)$ exhibit highly oscillatory behavior.
In such situations, approximation by polynomials of finite degree would not be appropriate.
To avoid these problems, one may need to choose a more suitable basis $|x\rangle$.

Let us mention one more related issue. 
The present approach relies crucially on the existence of an eigenvalue distribution. 
In the case of Minkowski-type matrix models, we extended this notion in a formal manner in~\eqref{eq:master field for m-type 1-m.m.}, where the existence of large-$N$ factorization served as the theoretical foundation. 
Based on perturbative analysis using ’t~Hooft graphs, as well as on the structure of the loop equations themselves, large-$N$ factorization is expected to hold for a broad class of matrix models, including Minkowski-type ones. 
However, there is no rigorous proof that it applies to all matrix models. 
Therefore, when applying the present method to other Minkowski-type matrix models, the validity of the approximation must be examined carefully. 
In particular, since the loop equations, viewed purely as a system of algebraic equations, also admit unphysical solutions, it is highly desirable to have a systematic and reliable consistency check to determine whether a given approximate solution indeed corresponds to a physical one. 

While these issues remain, we expect that it should in principle be possible to apply the present method to large-$N$ multi-matrix models, provided that they exhibit large-$N$–induced simplicity, as in the one-matrix model.
Further improvements and refinements of the computational method, as well as a more systematic theoretical understanding of large-$N$ factorization and master fields, will be required, but these issues are left for future work.

\acknowledgments

The author would like to thank Pei-Ming Ho, Tomohiro Inagaki, Hikaru Kawai, Henry Liao, Takeshi Morita, Jun Nishimura, and Hiromasa Watanabe for helpful discussions.
The author is also grateful to the organizers of the workshop “Thermal Quantum Field Theory and Their Applications” for their hospitality.
The author is supported by JST SPRING, Grant Number JPMJSP2132.

\appendix
\section{Exact Solution of the Euclidean-Type One-Matrix Model} \label{appendix a}

It has long been known that the Euclidean one-matrix model~\eqref{eq:S of one-matrix model} is exactly solvable, in the sense that expectation values of arbitrary observables can be computed exactly. Here we review one standard approach based on the eigenvalue distribution $\rho_{E}(\lambda)$ and the resolvent $R(z)$ (For more detailed calculation, for example, see~\cite{Eynard:2015aea}). The resolvent is defined by
\begin{align} \label{eq:resolvent}
R(z)\equiv\left\langle \frac{1}{N}\sum_{k=1}^{N}\frac{1}{z-\lambda_{k}}\right\rangle
=\left\langle \operatorname{tr}\!\left(\frac{1}{z-\phi}\right)\right\rangle,
\end{align}
which is a complex function that is holomorphic in the large-$N$ limit, except for a cut on the real axis. Expanding it around $|z|\to\infty$, one immediately finds from the geometric-series formula that
\begin{align} \label{eq:R as generating function}
R(z)=\sum_{n=0}^{\infty}\frac{w_{n}}{z^{n+1}}.
\end{align}
Thus $R(z)$ serves as a generating function for the moments $w_{n}$, and $w_{n}$ can be extracted by a counterclockwise contour integral along an infinitely large circle enclosing the origin,
\begin{align} \label{eq:residue theorem for w_n}
w_{n}=\oint_{C}\frac{dz}{2\pi i}\, z^{n}R(z).
\end{align}
As noted above, since $R(z)$ is holomorphic except for the cut on the real axis, this contour integral can be deformed, by Cauchy's theorem, to a contour encircling the cut. This is equivalent to integrating the discontinuity of $R(z)$ across the cut. Therefore, defining the eigenvalue distribution by
\begin{align} \label{eq:discontinuity and ED}
\rho_{E}(\lambda)\equiv\frac{1}{2\pi i}\bigl(R(\lambda-i\epsilon)-R(\lambda+i\epsilon)\bigr),
\end{align}
one recovers the familiar relation $w_{n}=\int_{\Omega} d\lambda\, \lambda^{n}\rho_{E}(\lambda)$. Here the integration domain $\Omega$ is precisely the cut of $R(z)$ on the real axis. Since $R(\lambda-i\epsilon)-R(\lambda+i\epsilon)=0$ outside $\Omega$, the density $\rho_{E}(\lambda)$ can be nonzero only on $\Omega$, and hence one may extend the integration domain to the entire real axis without changing the result.

Since the resolvent $R(z)$ contains the complete information about the eigenvalue distribution $\rho_{E}(\lambda)$, once the exact form of $R(z)$ is known, $\rho_{E}(\lambda)$ and hence all moments $w_{n}$ can be determined. It is well known that $R(z)$ can be obtained from the Schwinger--Dyson equations (by convention, this Schwinger--Dyson equation for $R(z)$ is also called the ``loop equation''). The starting point is the identity
\begin{align} \label{eq:derivation of loop eq about R}
\int\Bigl(\prod_{i=1}^{N} d\lambda_{i}\Bigr)\,
\frac{\partial}{\partial\lambda_{k}}
\left\{
\frac{1}{z-\lambda_{k}}\,
\Delta^{2}(\lambda)\,
e^{-\sum_{i=1}^{N}N V(\lambda_{i})}
\right\}=0,
\end{align}
where $V(\lambda_i)$ denotes the potential written in terms of the eigenvalues
$\lambda_1,\ldots,\lambda_N$ obtained by diagonalizing $\phi$,\footnote{For the moment, we assume that $V(\lambda)$ is a polynomial in $\lambda$.}
and $\Delta(\lambda)$ is the Vandermonde determinant,
$\Delta(\lambda)=\prod_{i<j}(\lambda_i-\lambda_j)$. Since the factor $\Delta^2(\lambda)$ can be regarded as a kind of Faddeev--Popov determinant, under this gauge fixing, expectation values are computed as
\begin{align}
\langle f(\lambda)\rangle
=\frac1Z \int\Bigl(\prod_{i=1}^{N} d\lambda_i\Bigr)\,
f(\lambda)\,\Delta^2(\lambda)\,e^{-\sum_{i=1}^{N}N V(\lambda_{i})} ,
\end{align}
the previous identity is reduced to the quadratic equation
\begin{align}
R^{2}(z)-(z-gz^{3})R(z)-g(z^{2}+zw_{1}+w_{2})+1=0.
\end{align}
Applying the quadratic formula, we obtain
\begin{align}
R(z)=\frac{1}{2}\Bigl[z-gz^{3}
-\sqrt{(z-gz^{3})^{2}+4\{g(w_{2}+zw_{1}+z^{2})-1\}}\Bigr].
\end{align}
The sign of the square root is chosen so that $R(z)\to 1/z$ as $|z|\to\infty$.
As is evident, $R(z)$ has at most six poles on the complex plane and therefore exhibits the associated discontinuities.
These discontinuities define the eigenvalue distribution $\rho_{E}(\lambda)$ shown in~\eqref{eq:discontinuity and ED}.

The resolvent $R(z)$ derived above still contains $w_{1}$ and $w_{2}$. Since these are determined by the eigenvalue distribution $\rho_{E}(\lambda)$, one could in principle determine $w_{1}$, $w_{2}$, and $\rho_{E}(\lambda)$ self-consistently. In practice, however, it is convenient to use the ``single-cut ansatz'' to extract $\rho_{E}(\lambda)$ from $R(z)$, and this also makes the connection to (and the difference from) the Minkowski-type one-matrix model more transparent. The procedure is as follows. First, we assume the $\mathbb{Z}_{2}$ symmetry of the theory (including quantum corrections), so that $w_{1}=0$. Next, we impose the following ``single-cut solution'' ansatz:
\begin{align}
\sqrt{(z-gz^{3})^{2}+4\{g(w_{2}+z^{2})-1\}}
\overset{!}{=} g(z-b)(z-c)\sqrt{(z-a)(z+a)},\qquad a\in\mathbb{R}.
\end{align}
This ansatz means that the expression under the square root in $R(z)$ takes the form
$(z-a)(z+a)(z-b)^{2}(z-c)^{2}$,
so that the would-be six branch points degenerate and the number of distinct branch points is reduced to two. Since $(z-b)(z-c)$ lies outside the square root, the points $z=b,c$ are no longer branch points. The appearance of the two branch points at $z=\pm a$ follows from the assumed $\mathbb{Z}_{2}$ symmetry. Moreover, the requirement that $a$ be real is a physical one: the discontinuity of $R(z)$ defines the support of the eigenvalue distribution. Since the eigenvalues of a Hermitian one-matrix model must be real, the support of $\rho_{E}(\lambda)$ must lie on the real axis. If $\pm a$ were located away from the real axis, it would no longer be possible to place the cut of $R(z)$ on the real axis.

With the single-cut ansatz, the problem of determining $\rho_{E}(\lambda)$ and $w_{2}$ reduces to determining three parameters $a,b,c$. These can be fixed using the asymptotic expansion of $R(z)$ at $z\to\infty$ in~\eqref{eq:R as generating function}. First, one must arrange the positive powers of $z$ to cancel against $z-gz^{3}$ so that they vanish, which immediately implies $c=-b$. The remaining parameters $a$ and $b$ are then determined by requiring that the coefficients of $z^{1}$ and $z^{-1}$ become $0$ and $1$, respectively. As a result of the calculation, one finally obtains
\begin{equation} \label{eq:exact solution for Euclidean}
\begin{aligned}
\rho_{E}(\lambda) & =\frac{\bigl(1-\frac{a^{2}g}{2}-g\lambda^{2}\bigr)\sqrt{(a-\lambda)(a+\lambda)}}{2\pi},\\
w_{2} & =\frac{a^{4}-a^{6}g}{16},\\
a & =\sqrt{\frac{2\bigl(1-\sqrt{1-12g}\bigr)}{3g}},
\end{aligned}
\end{equation}
as the exact solution of the single-cut ansatz.

In addition to its remarkable simplicity, a major advantage of this method is that the critical value of the coupling constant, $g_{c}$, can be determined directly.
In the present case, the requirement that the eigenvalue distribution be supported on the real axis imposes the condition $a\in\mathbb{R}$, which in turn fixes the critical value to be $g_{c}=\frac{1}{12}$.
For $g>\frac{1}{12}$, this requirement is violated, indicating that the single-cut solution is no longer valid.

\section{Formal Solution of the Minkowski-Type One-Matrix Model} \label{appendix b}

It is fair to say that solution methods for the Euclidean one-matrix model are well established, including the approach based on the resolvent and the eigenvalue distribution reviewed in Appendix~\ref{appendix a}. In contrast, the Minkowski one-matrix model involves several subtle points. In this appendix we describe these subtleties and explain how we handle them in the present work.

Whether the theory is Euclidean or Minkowski, the relation~\eqref{eq:R as generating function} expressing that the resolvent is a generating function for the moments remains unchanged. Moreover, as far as the loop equation for $R(z)$ is concerned, the difference between the Euclidean and Minkowski cases arises only from the coefficient in front of the $V'(\lambda_{k})$ term in~\eqref{eq:derivation of loop eq about R}. Therefore, replacing the sign $-1$ by $i$ immediately yields the Minkowski loop equation. Concretely, the loop equation for the Minkowski one-matrix model is
\begin{align}
R^{2}(z)+i(z-gz^{3})R(z)+ig\bigl(z^{2}+zw_{1}+w_{2}\bigr)-i=0.
\end{align}

This equation can be derived exactly as a Schwinger--Dyson equation, and in principle one should be able to extract the moment data $w_n$ from the residue integral~\eqref{eq:residue theorem for w_n}, as in the Euclidean-type one-matrix model.
However, a subtlety arises when one attempts to infer the existence of an ``eigenvalue distribution'' from $R(z)$. Let us proceed as in the Euclidean case: solve the quadratic equation for $R(z)$, and impose the $\mathbb{Z}_{2}$ symmetry and a ``single-cut'' ansatz:
\begin{align}
R(z)
&=\frac{1}{2}\Bigl[-i(z-gz^{3})+ \sqrt{-(z-gz^{3})^{2}-4i\{g(w_{2}+z^{2})-1\}}\Bigr] \\
&\overset{!}{=} \frac{1}{2}\Bigl[-i(z-gz^{3})+ ig(z^{2}-\beta^{2})\sqrt{(z-\alpha)(z+\alpha)}\Bigr]. \label{eq:single cut for Minkowski} 
\end{align}
The first line is obtained by applying the quadratic formula and setting $w_{1}=0$, while the second line is the single-cut ansatz, and the two are required to be identically equal. The sign in front of the square root is chosen so that $R(z)\to 1/z$ as $z\to\infty$.

Having assumed the cut structure of $R(z)$ in this way, we should reconsider whether the assumption is justified in the first place. In the Euclidean model, $\phi$ is Hermitian and the Boltzmann weight $e^{-S}$ is real. Hence the eigenvalue distribution $\rho_{E}(\lambda)$ must have support on the real axis, which is why we imposed the constraint $a\in\mathbb{R}$. Furthermore, the assumption that $\rho_{E}(\lambda)$ is supported on a single connected interval---equivalently, that the square root has only one cut---implies a degeneracy of branch points: the sextic polynomial under the square root must take the form $(z^{2}-a^{2})(z-b)^{2}(z-c)^{2}$. Otherwise, in the residue integral~\eqref{eq:residue theorem for w_n}, multiple cuts would contribute to the moments $w_{n}$.

Turning to the Minkowski one-matrix model, however, the weight $e^{iS}$ is complex, and therefore the support of $\rho_{M}(z)$ is no longer constrained to lie on the real axis. In fact, it is not even clear whether it should lie on a straight line. Of course, by Cauchy's theorem one may deform an integration contour as long as no singularities are crossed, and hence the contour may be taken to be a curve or a line segment. Nevertheless, given that in the Euclidean theory the eigenvalues should lie on the real axis, this apparent arbitrariness is somewhat unsatisfactory. There is also another mathematical issue. In the Euclidean one-matrix model, one may formally define the eigenvalue distribution as
$\rho_{E}(\lambda)=\left\langle \frac{1}{N}\sum_{k=1}^{N}\delta(\lambda-\lambda_{k})\right\rangle$.
If one naively tries to apply this definition to the Minkowski case by complexifying the argument of the delta function and writing
\begin{align}
\rho_{M}(z)\overset{?}{=}\left\langle \frac{1}{N}\sum_{k=1}^{N}\delta(z-\lambda_{k})\right\rangle,
\end{align}
one immediately encounters a problem: a one-dimensional delta function $\delta(z)$ on the complex plane is not well defined in general. The well-defined object is instead $\delta^{2}(z)$, which behaves properly under area integrals on the complex plane. In other words, it is mathematically unclear whether an eigenvalue distribution analogous to the Euclidean one even exists in the Minkowski case, let alone whether it is well defined.

Even taking such subtle circumstances into account, it is at least true that the resolvent $R(z)$ is the generating function of moments $w_n$, namely that the integral of $R(z)$ along its cut yields $w_n$.  
Moreover, the assumption of a single-cut structure also works at least formally.  
We therefore temporarily accept~\eqref{eq:single cut for Minkowski} and proceed with the same calculation as in the Euclidean case.  
Using the fact that the asymptotic behavior of $R(z)$ for $z\to\infty$, one can obtain $\alpha$, $\beta$ and $w_2$ as
\begin{equation} \label{eq:formal solution of alpha}
\begin{aligned}
R(z) & = \frac{1}{2}\Bigl[-i(z-g z^{3}) - i\Bigl(g z^{2} + \frac{\alpha^{2} g}{2} - 1\Bigr)\sqrt{(z-\alpha)(z+\alpha)}\Bigr], \\
\alpha & = \sqrt{\frac{2\bigl(1 \pm i\sqrt{-1 + 12 i g}\bigr)}{3 g}}, \\
w_{2} & = \frac{- i \alpha^{4} + i \alpha^{6} g}{16} .
\end{aligned}
\end{equation}
Note that, for the $\pm$ sign inside the square root appearing in $\alpha$, one must choose the branch that does not diverge in the limit $g\to0$.
Here we fix $\sqrt{-1+12ig}$ to have a branch cut along the negative real axis $(-\infty,0]$, so that for $g\in\mathbb{R}$ the principal values are determined as
\begin{align}
\sqrt{-1+12ig}\underset{g\to0^{+}}{\longrightarrow}+i,\quad
\sqrt{-1+12ig}\underset{g\to0^{-}}{\longrightarrow}-i.
\end{align}
Therefore, the plus sign must be chosen for $g>0$, while the minus sign must be chosen for $g<0$.
With these points in mind, the power-series expansion of $w_{2}$ is independent of the sign of $g$ and takes the form
\begin{align}
w_{2} = i - 2 g - 9 i g^{2} + \cdots .
\end{align}
In fact, this exactly coincides with the result obtained from the formal power-series expansion around $g=0$.
Thus, at least for $|g|\ll1$, one can see that even in the Minkowski model the assumption of a one-cut structure reproduces perturbation theory.\footnote{This fact was pointed out to us by Hikaru Kawai and Henry Liao.}

Furthermore, as emphasized repeatedly, the support of the present ``eigenvalue distribution'' does not need to lie on the real axis, and therefore $\alpha$ is allowed to be complex.  
This leads to a major difference from the Euclidean model.  
In the Euclidean case, the requirement that $a$ be real leads to the critical value $g_c = \frac{1}{12}$.  
For $g$ exceeding this value, the single-cut assumption breaks down and the theory undergoes a phase transition (in fact, for $g > g_c$ the matrix integral itself is no longer well defined).  
Presumably reflecting this fact, the least-squares analysis in Section~\ref{sec:Numerical Results} fails to converge in the region $g > g_c$.  
This strongly suggests that the two self-consistency conditions of the present bootstrap approximation, namely that an eigenvalue distribution exists such that $w_n = \int d\lambda\, \lambda^n \rho(\lambda)$ and that the moments $w_n$ satisfy the loop equations, can no longer be simultaneously satisfied.  

In the Minkowski model, by contrast, $\alpha$ is no longer constrained to be real.
Therefore, at least the same mechanism of phase transition as in the Euclidean case does not seem to occur.
However, one must note that the one-cut assumption that serves as the starting point of the present discussion is not supported by a clear theoretical justification as in the Euclidean case, but rather should be regarded as a working hypothesis.
If this one-cut assumption were to fail, the arguments concerning the phase transition would necessarily have to be reconsidered.
At present it remains an open question whether a critical value $g_{c}$ exists in the Minkowski one-matrix model.
We also expect that the instability observed around $g\sim0$ in the numerical calculation is most likely a numerical artifact arising from the particular choice of the cut described above.


 \bibliographystyle{JHEP}
 \bibliography{biblio}


\end{document}